\documentclass[aps,prd,twocolumn,superscriptaddress,nofootinbib]{revtex4-1}

\usepackage{latexsym}
\usepackage{amsmath}
\usepackage{amssymb}
\usepackage{amsfonts}
\usepackage{bm}

\usepackage{color}
\definecolor{purple}{rgb}{0.5,0,0.5}
\definecolor{blue}{rgb}{0.0,0,0.9}
\definecolor{prdblue}{rgb}{0.133,0.118,0.498}
\usepackage[colorlinks=true, pdfstartview=FitV, linkcolor=prdblue, citecolor= prdblue, urlcolor=prdblue]{hyperref}

\usepackage{supertabular} 
\usepackage{placeins}
\usepackage{epsfig}
\usepackage{graphicx}

\begin{document}


\title{Fully strange tetra- and penta-quarks in a chiral quark model}


\author{Gang Yang}
\email[]{yanggang@zjnu.edu.cn}
\affiliation{Department of Physics, Zhejiang Normal University, Jinhua 321004, China}

\author{Jialun Ping}
\email[]{jlping@njnu.edu.cn}
\affiliation{Department of Physics and Jiangsu Key Laboratory for Numerical Simulation of Large Scale Complex Systems, Nanjing Normal University, Nanjing 210023, P. R. China}

\author{Jorge Segovia}
\email[]{jsegovia@upo.es}
\affiliation{Departamento de Sistemas F\'isicos, Qu\'imicos y Naturales, Universidad Pablo de Olavide, E-41013 Sevilla, Spain}



\begin{abstract}
Motivated by the recently reported resonant structure $X(2300)$, a strong candidate for a fully strange tetraquark with positive parity, we perform a systematic study of fully strange tetra- and penta-quark systems within a chiral quark model. Low-lying $S$-wave configurations of the $ss\bar s\bar s$ and $ssss\bar s$ systems are investigated using the Gaussian Expansion Method (GEM) combined with the Complex Scaling Method (CSM), which allows for a unified treatment of bound, resonant, and scattering states. For tetraquarks, all possible configurations: meson-meson, diquark-antidiquark, and K-type structures, with complete color bases, are incorporated, while baryon-meson and diquark-diquark-antiquark configurations are considered for pentaquarks. Several weakly bound states and narrow resonances are identified in both sectors. In particular, a compact fully strange tetraquark with $J^P=1^+$ is found near $2.3\,\text{GeV}$, providing a natural interpretation of the $X(2300)$ resonance. Additional exotic states with dominant hidden-color and K-type components are predicted in the mass ranges $1.6-3.1$ GeV for tetraquarks and $2.6-3.2$ GeV for pentaquarks. The internal structure of these states is analyzed through their sizes, magnetic moments, and wave-function compositions, highlighting the essential role of channel coupling and exotic color configurations. Finally, promising two-body strong decay channels are proposed to facilitate future experimental searches.
\end{abstract}

\pacs{
12.38.-t \and 
12.39.-x      
}
\keywords{
Quantum Chromodynamics \and
Quark models
}

\maketitle


\section{Introduction}
\label{sec:intro}

The exploration of exotic hadrons beyond the conventional quark-antiquark mesons and three-quark baryons has become one of the central topics in contemporary hadron spectroscopy. In recent years, numerous candidates for multiquark states have been reported experimentally, stimulating extensive theoretical efforts aimed at understanding their internal structure and underlying QCD dynamics. Among these systems, hadrons composed exclusively of strange quarks and antiquarks occupy a special place: being still considered light, they are in the intersection of light and heavy quark sectors, offering an alternative laboratory to investigate multiquark dynamics driven by nonperturbative QCD.

Recently, the BESIII Collaboration reported a resonant structure, denoted as $X(2300)$, in a partial-wave analysis of the process $\psi(3683)\rightarrow \phi \eta \eta'$~\cite{BESIII:2025wpp}. This state is particularly intriguing, as it represents a strong candidate for the first fully strange tetraquark with positive parity. If confirmed, such a state would significantly enrich our understanding of exotic hadrons in the light-flavor sector and provide valuable constraints on theoretical models of multiquark confinement.

Over the past two decades, several structures interpreted as fully strange exotic states have been observed experimentally, albeit predominantly with negative parity. The $\phi(2170)$, first reported by the BaBar Collaboration~\cite{BaBar:2006gsq, BaBar:2007ptr, BaBar:2007ceh, BaBar:2011btv} and subsequently confirmed by BES, BESIII~\cite{BES:2007sqy, BESIII:2014ybv, BESIII:2017qkh}, and Belle~\cite{Belle:2008kuo}, has attracted sustained attention. Additional structures, including the $X(2100)$~\cite{BESIII:2018zbm}, $X(2239)$~\cite{BESIII:2018ldc}, $X(2370)$~\cite{BESIII:2010gmv, BESIII:2019wkp}, and $X(2500)$~\cite{BESIII:2016qzq}, were reported by BES and BESIII in various decay channels. These observations have triggered a broad range of theoretical interpretations.

From the theoretical side, the $\phi(2170)$ has been interpreted as a fully strange tetraquark with $J^{PC}=1^{--}$ within QCD sum rules~\cite{Wang:2006ri, Chen:2008ej, Chen:2018kuu, Jiang:2023atq} and constituent quark models~\cite{Deng:2010zzd, Drenska:2008gr}. Alternative scenarios have also been proposed, including strangeonium hybrid states~\cite{Ding:2006ya, Li:2025hsp, Ho:2019org}, excited $s\bar s$ mesons~\cite{Ding:2007pc}, hadronic molecules~\cite{MartinezTorres:2008gy, Alvarez-Ruso:2009vkn, Dong:2017rmg}, and diquark-antidiquark configurations~\cite{Agaev:2019coa}. The $X(2100)$ and $X(2370)$ have been associated with $0^{-+}$~\cite{Dong:2020okt, Su:2022eun} and $1^{+-}$~\cite{Su:2022eun, Cui:2019roq, Wang:2019nln, Azizi:2019ecm, Cao:2024mfn} fully strange tetraquark states, respectively, while the $X(2500)$ has been suggested as a $P$-wave $0^-+$ state in both QCD sum rules and quark-model studies~\cite{Dong:2020okt, Su:2022eun, Liu:2020lpw, Lu:2019ira}.

In contrast, fully strange tetraquarks with positive parity have received comparatively less attention. For the recently observed $X(2300)$, QCD sum-rule calculations favor an interpretation as an $ss\bar s\bar s$ tetraquark with $J^P=1^+$~\cite{Wan:2025xhf}, while effective models analyzing its production and decay mechanisms also support an $S$-wave multiquark configuration~\cite{Cao:2025dze, Liu:2026ljb}. More generally, several exotic fully strange states in the mass range $2.1-3.6$ GeV have been predicted using QCD sum rules and quark models~\cite{Xi:2023byo, Wang:2024pgy, Wu:2025mae, Ma:2024vsi}. Nevertheless, important questions remain open: Which configurations dominate the low-lying fully strange spectrum? What is the role of hidden-color and exotic spatial arrangements? Can bound and resonant states be unambiguously distinguished from continuum effects?

Motivated by these open issues, we present a systematic and unified study of fully strange tetra- and penta-quark systems within a chiral constituent quark model. Our analysis focuses on the low-lying S-wave $ss\bar s\bar s$ and $ssss\bar s$ configurations, employing the Gaussian Expansion Method (GEM) in conjunction with the Complex Scaling Method (CSM). This framework allows us to treat bound states, resonances, and scattering states on the same footing, and has demonstrated considerable success in the study of a wide variety of exotic hadrons, including hidden-, single-, double-, triple-, and fully heavy tetraquarks~\cite{gy:2020dht, gy:2020dhts, Yang:2021hrb, Yang:2023mov, Yang:2021zhe, Yang:2022cut, Yang:2021izl, Yang:2023mov, Yang:2023wgu, Yang:2024nyc, Yang:2025jsp}, as well as several classes of pentaquarks~\cite{Yang:2015bmv, Yang:2018oqd, gy:2020dcp, Yang:2020twg, Yang:2022bfu, Yang:2023dzb}. 
 
A distinctive feature of the present work is the complete inclusion of all $S$-wave configurations and color structures relevant to fully strange multiquark systems. For tetraquarks, we consider meson-meson, diquark-antidiquark, and K-type configurations, incorporating both color-singlet and hidden-color channels. The latter, and in particular the K-type structures, are often neglected or treated incompletely in previous studies, despite their potential importance in multiquark dynamics. For pentaquarks, both baryon-meson and diquark-diquark-antiquark configurations are simultaneously included within all allowed color couplings.

Beyond the determination of mass spectra, we investigate the internal structure of the resulting exotic states by analyzing their root-mean-square radii, magnetic moments, and wave-function compositions. These observables provide valuable insight into whether a given state exhibits a compact multiquark nature or is dominated by extended molecular components. In addition, we identify dominant decay channels for selected bound and resonant states, thereby offering concrete predictions that can be tested in future experimental searches.

This paper is organized as follows. In Section~\ref{sec:model}, we introduce the chiral quark model and outline the construction of the tetra- and penta-quark wave functions. Section~\ref{sec:results} presents our numerical results and a detailed discussion of the bound and resonant fully strange multiquark states. Finally, Section~\ref{sec:summary} summarizes our main conclusions and highlights perspectives for future studies.


\section{Theoretical framework}
\label{sec:model}

\subsection{The Hamiltonian}

To investigate possible four- and five-body bound and resonant states, we employ the complex-scaled Schr\"odinger equation,
\begin{equation}\label{CSMSE}
\left[ H(\theta)-E(\theta) \right] \Psi_{JM}(\theta)=0 \,,
\end{equation}
where $\theta$ denotes the complex-scaling angle.

Within a QCD-inspired chiral quark model, the Hamiltonian of an $n$-body system is written as
\begin{equation}
H(\theta) = \sum_{i=1}^{n} \left( m_i + \frac{(\vec{p}_i\, e^{-i\theta})^2}{2m_i}\right) - T_{\text{CM}} + \sum_{j>i=1}^{n} V(\vec{r}_{ij} e^{i\theta}) \,,
\label{eq:Hamiltonian}
\end{equation}
where $m_i$ and $\vec{p}_i$ denote the constituent mass and momentum of the $i$th quark, respectively, $T_{\text{CM}}$ is the center-of-mass kinetic energy, and $V$ represents the two-body interaction potential.

The complex scaling is implemented by introducing a rotation angle $\theta$ into both the kinetic and potential terms. Within this framework, the eigenvalues of Eq.~\eqref{CSMSE} naturally separate into three categories: bound states, resonant states, and scattering states. Bound states appear as real eigenvalues on the energy axis and are independent of $\theta$. Resonant states are identified as $\theta$-independent complex eigenvalues located above the corresponding threshold lines, with decay widths given by $\Gamma=-2\,\text{Im}(E)$. Scattering states, by contrast, are distributed along the rotated branch cuts and vary with the scaling angle.

The dynamics of fully strange tetra- and penta-quark systems, $ss\bar s\bar s$ and $ssss\bar{s}$, are governed by the two-body complex-scaled potential
\begin{equation}
\label{CQMV}
V(\vec{r}_{ij} e^{i\theta}) = V_{\text{CON}}(\vec{r}_{ij} e^{i\theta}) + V_{\text{OGE}}(\vec{r}_{ij} e^{i\theta}) + V_{\chi}(\vec{r}_{ij} e^{i\theta}) \,.
\end{equation}
Here, $V_{\text{CON}}$, $V_{\text{OGE}}$, and $V_{\chi}$ denote the color confinement, one-gluon exchange, and Goldstone-boson exchange interactions, respectively, which capture the essential nonperturbative and perturbative features of QCD. Since the present study focuses on low-lying $S$-wave multiquark states, only the central and spin-spin components of these interactions are considered.

The color confinement interaction is motivated by lattice QCD results, which indicate that multi-gluon exchanges generate a linearly rising potential at short distances, while string breaking due to light-quark pair creation leads to saturation at larger distances~\cite{Bali:2005fu}. These features are phenomenologically incorporated through
\begin{equation}
V_{\text{CON}}(\vec{r}_{ij} e^{i\theta})=\left[-a_{c}(1-e^{-\mu_{c}r_{ij} e^{i\theta}})+\Delta \right] (\lambda_{i}^{c}\cdot \lambda_{j}^{c}) \,,
\label{eq:conf}
\end{equation}
where $\lambda^{c}$ are the SU(3) color Gell-Mann matrices, and $a_{c}$, $\mu_{c}$, and $\Delta$ are model parameters. For $\theta=0^\circ$, this potential behaves linearly at short distances with an effective confinement strength $\sigma=-a_{c}\mu_{c}(\lambda^{c}_{i}\cdot\lambda^{c}_{j})$, while it saturates to a constant value $V_{\text{thr.}} = (\Delta-a_{c}) (\lambda^{c}_{i}\cdot\lambda^{c}_{j})$ at large separations.

At energy scales above chiral symmetry breaking, perturbative QCD effects become relevant. The leading contribution is provided by the one-gluon exchange interaction, which includes both Coulomb and color-magnetic terms,
\begin{align}
V_{\text{OGE}}(\vec{r}_{ij} e^{i\theta}) &= \frac{1}{4} \alpha_{s} (\lambda_{i}^{c}\cdot \lambda_{j}^{c}) \Bigg[ \frac{1}{r_{ij} e^{i\theta}} \nonumber \\
& 
- \frac{1}{6m_{i}m_{j}} (\vec{\sigma}_{i}\cdot\vec{\sigma}_{j})
\frac{e^{-r_{ij} e^{i\theta} /r_{0}(\mu_{ij})}}{r_{ij} e^{i\theta} r_{0}^{2}(\mu_{ij})} \Bigg] \,,
\end{align}
where $\vec{\sigma}$ are the Pauli matrices. The smearing parameter $r_0(\mu_{ij})=\hat{r}_0/\mu_{ij}$ depends on the reduced mass $\mu_{ij}$ of the interacting quark pair. The regularized contact term corresponds to
\begin{equation}
\delta(\vec{r}_{ij} e^{i\theta}) \sim \frac{1}{4\pi r_{0}^{2}(\mu_{ij})}
\frac{e^{-r_{ij} e^{i\theta}/r_{0}(\mu_{ij})}}{r_{ij} e^{i\theta}} \,.
\end{equation}

The effective strong coupling constant is taken to be scale dependent and frozen at low energies, following Ref.~\cite{Segovia:2013wma},
\begin{equation}
\alpha_{s}(\mu_{ij})=\frac{\alpha_{0}}{\ln\left(\frac{\mu_{ij}^{2}+\mu_{0}^{2}}{\Lambda_{0}^{2}} \right)} \,,
\end{equation}
where $\alpha_{0}$, $\mu_{0}$, and $\Lambda_{0}$ are model parameters.

For fully strange systems, the relevant Goldstone-boson exchange interactions arise from $\sigma$ and $\eta$ meson exchange. Their central contributions are given by
\begin{align}
V_{\sigma}\left( \vec{r}_{ij} e^{i\theta} \right) &= - \frac{g_{ch}^{2}}{4\pi}
\frac{\Lambda_{\sigma}^{2}}{\Lambda_{\sigma}^{2}-m_{\sigma}^{2}}m_{\sigma} \nonumber \\
&
\times \Bigg[ Y(m_{\sigma}r_{ij}e^{i\theta}) - \frac{\Lambda_{\sigma}}{m_{\sigma}}Y(\Lambda_{\sigma}r_{ij}e^{i\theta}) \Bigg] \,, \\
V_{\eta}\left( \vec{r}_{ij} e^{i\theta}\right) &= \frac{g_{ch}^{2}}{4\pi}
\frac{m_{\eta}^2}{12m_{i}m_{j}} \frac{\Lambda_{\eta}^{2}}{\Lambda_{\eta}^{2}-m_{\eta}^{2}}m_{\eta} \nonumber \\
&
\times \Bigg[ Y(m_{\eta}r_{ij}e^{i\theta}) - \frac{\Lambda_{\eta}^{3}}{m_{\eta}^{3}}Y(\Lambda_{\eta}r_{ij}e^{i\theta}) \Bigg]  \nonumber \\
&
\times (\vec{\sigma}_{i}\cdot\vec{\sigma}_{j}) \Big[ \cos\theta_{p}(\lambda_{i}^{8}\cdot\lambda_{j}^{8}) - \sin\theta_{p} \Big] \,,
\end{align}
where $Y(x)=e^{-x}/x$ is the Yukawa function, $\lambda^{8}$ is the SU(3) flavor Gell-Mann matrix, and $\theta_{p}$ is the pseudoscalar mixing angle used to describe the physical $\eta$ meson. The $\eta$ mass is taken from experiment, while the $\sigma$ mass is determined via the PCAC relation $m_{\sigma}^{2}\simeq m_{\pi}^{2}+4m_{u,d}^{2}$~\cite{Scadron:1982eg}. The chiral coupling constant $g_{ch}$ is fixed through the $\pi NN$ coupling,
\begin{equation}
\frac{g_{ch}^{2}}{4\pi}=\frac{9}{25}\frac{g_{\pi NN}^{2}}{4\pi}
\frac{m_{u,d}^{2}}{m_{N}^2} \,,
\end{equation}
assuming SU(3) flavor symmetry broken only by the strange quark mass.

All model parameters are summarized in Table~\ref{tab:model} and are taken from Ref.~\cite{Vijande:2004he}. This chiral quark model has been successfully applied to describe hadron spectra, hadron-hadron interactions, and multiquark systems across a wide range of energies~\cite{Fernandez:1993hx, Valcarce:1994nr, Valcarce:1995dm, Vijande:2006jf, Segovia:2008zza, Segovia:2008zz, Segovia:2009zz, Ortega:2009hj, Segovia:2011zza, Segovia:2015dia, Yang:2015bmv, Ortega:2016hde, Ortega:2016mms, Ortega:2016pgg, Yang:2017xpp, Yang:2017rpg, Ortega:2018cnm, Ortega:2020uvc, Yang:2020atz}.

Finally, in order to clearly identify continuum thresholds associated with meson-meson or baryon-meson configurations in the complex energy plane, the theoretical and experimental masses of the ground and first radially excited $s\bar{s}$ mesons and the $\Omega$ baryon are listed in Table~\ref{MesonMass}. The theoretical values are calculated within the present framework, while the experimental masses are taken from Ref.~\cite{ParticleDataGroup:2024cfk}.

\begin{table}[!t]
	\caption{\label{tab:model} The chiral quark model parameters taken from Ref.~\cite{Vijande:2004he}.}
	\begin{ruledtabular}
		\begin{tabular}{llr}
			Quark mass & $m_s$ (MeV) &  555 \\[2ex]
			Confinement      & $a_c$ (MeV)         & 430 \\
			& $\mu_c$ (fm$^{-1})$ & 0.70 \\
			& $\Delta$ (MeV)      & 181.10 \\[2ex]
			OGE          & $\alpha_0$              & 2.118 \\
			& $\Lambda_0~$(fm$^{-1}$) & 0.113 \\
			& $\mu_0~$(MeV)           & 36.976 \\
			& $\hat{r}_0~$(MeV~fm)    & 28.17 \\[2ex]
			Goldstone bosons & $\Lambda_\sigma~$ (fm$^{-1}$) &   4.20 \\
			& $\Lambda_\eta$ (fm$^{-1}$)     &   5.20 \\
			& $g^2_{ch}/(4\pi)$                         &   0.54 \\
			& $\theta_P(^\circ)$                        & -15 \\
		\end{tabular}
	\end{ruledtabular}
\end{table}

\begin{table}[!t]
	\caption{\label{MesonMass} Masses ($M_{\text{The.}}$), in MeV, of $1S$ and $2S$ states of $s\bar{s}$ mesons and $\Omega$ baryon are calculated within the chiral quark model. Experimental mass ($M_{\text{Exp.}}$) values are taken from Ref.~\cite{ParticleDataGroup:2024cfk}.}
	\begin{ruledtabular}
		\begin{tabular}{crrr}
			Hadron & $nL$ & $M_{\text{The.}}$ & $M_{\text{Exp.}}$\\
			\hline
			$\phi$ & $1S$ &  $1011$ & $1020$ \\
			& $2S$ & $1720$ & $1680$        \\[2ex]
			$\eta'$ & $1S$ &  $828$ & $958$ \\
			& $2S$ & $1639$ & $1760$       \\[2ex]
			$\Omega$ & $1S$ &  $1634$ & $1672$ \\
			& $2S$ & $2226$ & $-$
		\end{tabular}
	\end{ruledtabular}
\end{table}


\begin{figure}[ht]
	\epsfxsize=3.4in \epsfbox{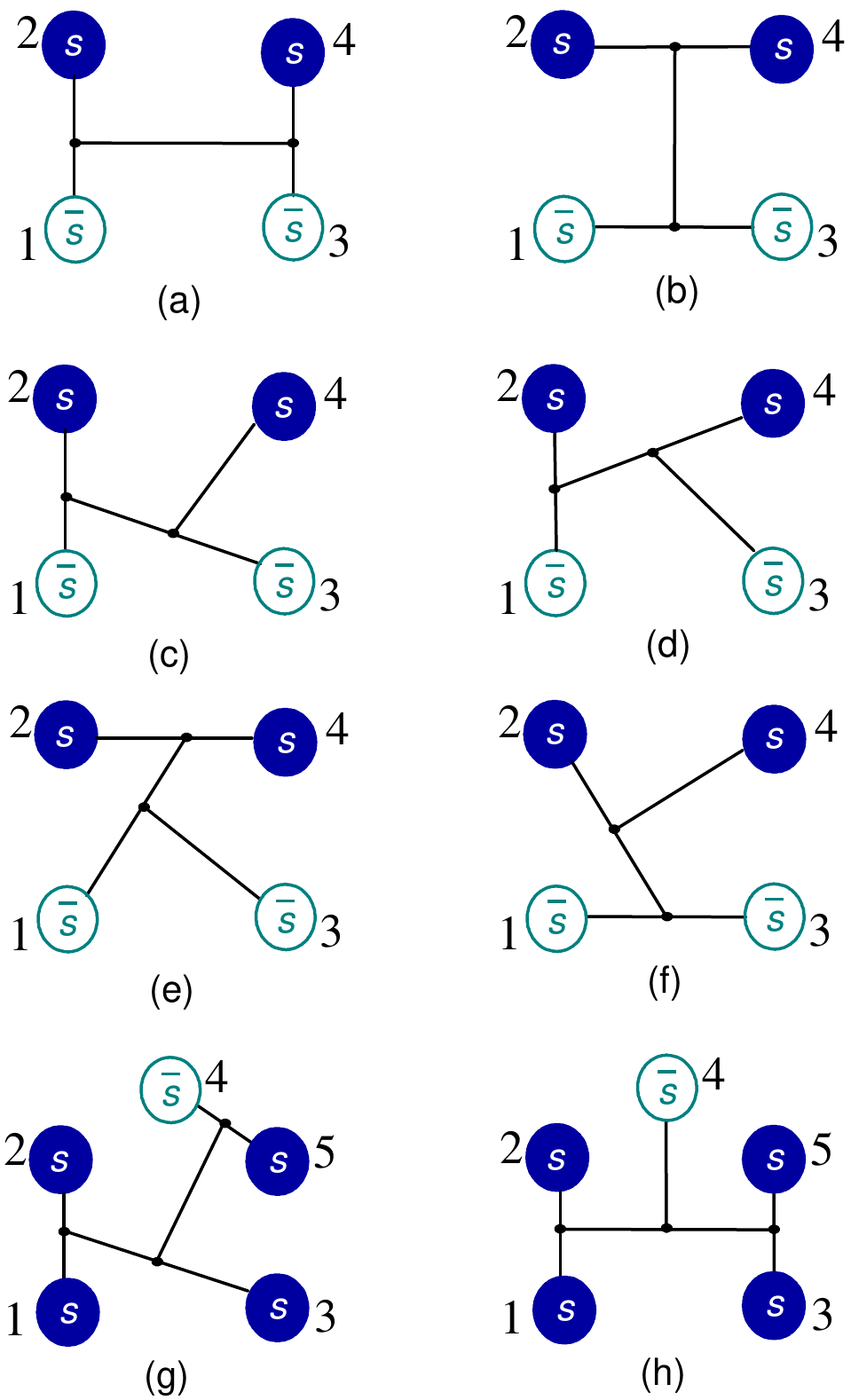}
	\caption{\label{QQqq} The $S$-wave configurations considered in this investigation for the fully strange tetra- and penta-quarks. Particularly, panel $(a)$ is a meson-meson structure, panel $(b)$ is a diquark-antidiquark arrangement, and four K-type configurations are from panel $(c)$ to $(f)$. Panel $(g)$ and $(h)$ is a baryon-meson and diquark-diquark-antiquark structure of pentaquark, respectively.}
\end{figure}

\subsection{The tetraquark wave function}

Figure~\ref{QQqq} illustrates the $S$-wave configurations of fully strange tetra- and penta-quark systems considered in this work. For the $ss\bar s\bar s$ tetraquark, all possible $S$-wave structures are included, while two distinct configurations are considered for the $ssss\bar s$ pentaquark. Specifically, the tetraquark sector consists of one meson-meson configuration [Fig.~\ref{QQqq}(a)], one diquark-antidiquark configuration [Fig.~\ref{QQqq}(b)], and four K-type configurations [Fig.~\ref{QQqq}(c)--(f)], which are rarely discussed in existing studies of multiquark systems. For the pentaquark case, Fig.~\ref{QQqq}(g) and (h) correspond to baryon-meson and diquark-diquark-antiquark configurations, respectively. The inclusion of this diverse set of configurations provides an efficient and comprehensive framework for the investigation of fully strange multiquark states.

The total wave function of a hadron is constructed as the direct product of its color, spin, flavor, and spatial components at the quark level. We begin with the color degree of freedom. For the $ss\bar{s}\bar{s}$ tetraquark in the meson-meson configuration, a color-singlet wave function can be obtained by coupling two color-singlet clusters, $1\otimes1$,
\begin{align}
\label{Color1}
\chi^c_1 &= \frac{1}{3}(\bar{r}r+\bar{g}g+\bar{b}b)\times (\bar{r}r+\bar{g}g+\bar{b}b) \,,
\end{align}
or, alternatively, by coupling two color-octet clusters, $8\otimes8$,
\begin{align}
\label{Color2}
\chi^c_2 &= \frac{\sqrt{2}}{12}(3\bar{b}r\bar{r}b+3\bar{g}r\bar{r}g+3\bar{b}g\bar{g}b+3\bar{g}b\bar{b}g+3\bar{r}g\bar{g}r
\nonumber\\
&+3\bar{r}b\bar{b}r+2\bar{r}r\bar{r}r+2\bar{g}g\bar{g}g+2\bar{b}b\bar{b}b-\bar{r}r\bar{g}g
\nonumber\\
&-\bar{g}g\bar{r}r-\bar{b}b\bar{g}g-\bar{b}b\bar{r}r-\bar{g}g\bar{b}b-\bar{r}r\bar{b}b) \,.
\end{align}
The former corresponds to the conventional color-singlet channel, while the latter represents a hidden-color configuration.

The color wave functions associated to the diquark-antidiquark arrangement are the coupled color triplet-antitriplet clusters, $3\otimes \bar{3}$:
\begin{align}
\label{Color3}
\chi^c_3 &= \frac{\sqrt{3}}{6}(\bar{r}r\bar{g}g-\bar{g}r\bar{r}g+\bar{g}g\bar{r}r-\bar{r}g\bar{g}r+\bar{r}r\bar{b}b
\nonumber\\
&-\bar{b}r\bar{r}b+\bar{b}b\bar{r}r-\bar{r}b\bar{b}r+\bar{g}g\bar{b}b-\bar{b}g\bar{g}b
\nonumber\\
&+\bar{b}b\bar{g}g-\bar{g}b\bar{b}g) \,,
\end{align}
and the coupled color sextet-antisextet clusters, $6\otimes \bar{6}$:
\begin{align}
\label{Color4}
\chi^c_4 &= \frac{\sqrt{6}}{12}(2\bar{r}r\bar{r}r+2\bar{g}g\bar{g}g+2\bar{b}b\bar{b}b+\bar{r}r\bar{g}g+\bar{g}r\bar{r}g
\nonumber\\
&+\bar{g}g\bar{r}r+\bar{r}g\bar{g}r+\bar{r}r\bar{b}b+\bar{b}r\bar{r}b+\bar{b}b\bar{r}r
\nonumber\\
&+\bar{r}b\bar{b}r+\bar{g}g\bar{b}b+\bar{b}g\bar{g}b+\bar{b}b\bar{g}g+\bar{g}b\bar{b}g) \,.
\end{align}

For the K-type configurations shown in Fig.~\ref{QQqq}(c)--(f), the associated color wave functions are constructed analogously. The color wave functions $\chi^c_5$ and $\chi^c_6$, corresponding to the $K_1$ configuration [Fig.~\ref{QQqq}(c)], are given by 
\begin{align}
\chi^c_5 = \chi^c_2,\ \ \ \ \chi^c_6 =\chi^c_1.
\end{align}
The color wave functions $\chi^c_7$ and $\chi^c_8$ from the $K_2$ configuration [Fig.~\ref{QQqq}(d)] are similar and they are
\begin{align}
\chi^c_7 = \chi^c_1,\ \ \ \ \chi^c_8 =\chi^c_2.
\end{align}
The color wave functions $\chi^c_9$ and $\chi^c_{10}$ from the $K_3$ configuration [Fig.~\ref{QQqq}(e)] read
\begin{eqnarray}
\chi^c_9 &=& \frac{1}{2\sqrt{6}}( \bar{r}b\bar{b}r+\bar{r}r\bar{b}b+\bar{g}b\bar{b}g+\bar{g}g\bar{b}b+\bar{r}g\bar{g}r+\bar{r}r\bar{g}g
\nonumber\\
&+& \bar{b}b\bar{g}g+\bar{b}g\bar{g}b+\bar{g}g\bar{r}r+\bar{g}r\bar{r}g+\bar{b}b\bar{r}r+\bar{b}r\bar{r}b)
\nonumber\\
&+& \frac{1}{\sqrt{6}}(\bar{r}r\bar{r}r+\bar{g}g\bar{g}g+\bar{b}b\bar{b}b),\nonumber\\
\chi^c_{10} &=& \frac{1}{2\sqrt{3}}( \bar{r}b\bar{b}r-\bar{r}r\bar{b}b+\bar{g}b\bar{b}g-\bar{g}g\bar{b}b+\bar{r}g\bar{g}r-\bar{r}r\bar{g}g
\nonumber\\
&-& \bar{b}b\bar{g}g+\bar{b}g\bar{g}b-\bar{g}g\bar{r}r+\bar{g}r\bar{r}g-\bar{b}b\bar{r}r+\bar{b}r\bar{r}b).
\end{eqnarray}
The color wave functions $\chi^c_{11}$ and $\chi^c_{12}$ from the $K_4$ configuration [Fig.~\ref{QQqq}(f)] are
\begin{align}
\chi^c_{11} = \chi^c_9,\ \ \ \ \chi^c_{12} =-\chi^c_{10}.
\end{align}

For the $ssss\bar{s}$ pentaquark system, the color-singlet and hidden-color bases of the baryon-meson configuration are given by 
\begin{align}
\label{Color1P}
\chi^{c}_{13} &= \frac{1}{\sqrt{18}}(rgb-rbg+gbr-grb+brg-bgr) \nonumber \\
&
\times (\bar r r+\bar gg+\bar bb) \,.
\end{align}
and
\begin{align}
\label{Color2P}
\chi^{c}_k &= \frac{1}{\sqrt{8}}(\chi^{k}_{3,1}\chi_{2,8}-\chi^{k}_{3,2}\chi_{2,7}-\chi^{k}_{3,3}\chi_{2,6}+\chi^{k}_{3,4}\chi_{2,5} \nonumber \\
& +\chi^{k}_{3,5}\chi_{2,4}-\chi^{k}_{3,6}\chi_{2,3}-\chi^{k}_{3,7}\chi_{2,2}+\chi^{k}_{3,8}\chi_{2,1}) \,,
\end{align}
respectively. Here, the index $k=14,15$ denotes symmetric and antisymmetric color configurations of the three-quark subcluster. The color bases employed are identical to those used in previous studies of hidden-charm, hidden-bottom, and doubly charmed pentaquark systems~\cite{Yang:2015bmv, Yang:2018oqd, gy:2020dcp}. For the diquark-diquark-antiquark configuration shown in Fig.~\ref{QQqq}(h), two independent color-singlet wave functions are constructed:
\begin{align}
\label{Color3P}
\chi^{c}_{16} &= \frac{1}{\sqrt{48}}\ \lbrace \bar{r}[(rb+br)(rg-gr)-(rg+gr)(rb-br)] \nonumber \\
&
+\bar{g}[(rg+gr)(gb-bg)+(gb+bg)(rg-gr)] \nonumber \\
&
+\bar{b}[(rb+br)(gb-bg)-(gb+bg)(rb-br)] \nonumber \\
&
+\sqrt{2}[\bar{r}rr(gb-bg)-\bar{g}gg(rb-br)+\bar{b}bb(rg-gr)] \rbrace\,,
\end{align}
and
\begin{align}
\label{Color4P}
\chi^{c}_{17} &= \frac{1}{\sqrt{24}}\ \lbrace \bar{r}[(rg-gr)(rb-br)-(rb-br)(rg-gr)] \nonumber \\
&
+\bar{g}[(rg-gr)(gb-bg)-(gb-bg)(rg-gr)] \nonumber \\
&
+\bar{b}[(rb-br)(gb-bg)-(gb-bg)(rb-br)] \rbrace\,.
\end{align}

Turning to the spin degree of freedom, the $S$-wave ground state of a tetraquark system can have total spin $S=0$, $1$ or $2$, while for pentaquarks the allowed total spin ranges from $1/2$ to $5/2$. The corresponding spin wave functions, denoted by $\chi^{\sigma_i}_{S M_S}$, are constructed for each configuration. Without loss of generality, the spin projection is chosen as $M_S=S$. For the four-body system, the explicit spin wave functions are given by
\begin{align}
\label{SWF0}
\chi_{0,0}^{\sigma_{u_1}}(4) &= \chi^\sigma_{00}\chi^\sigma_{00} \,, \\
\chi_{0,0}^{\sigma_{u_2}}(4) &= \frac{1}{\sqrt{3}}(\chi^\sigma_{11}\chi^\sigma_{1,-1}-\chi^\sigma_{10}\chi^\sigma_{10}+\chi^\sigma_{1,-1}\chi^\sigma_{11}) \,, \\
\chi_{0,0}^{\sigma_{u_3}}(4) &= \frac{1}{\sqrt{2}}\big((\sqrt{\frac{2}{3}}\chi^\sigma_{11}\chi^\sigma_{\frac{1}{2}, -\frac{1}{2}}-\sqrt{\frac{1}{3}}\chi^\sigma_{10}\chi^\sigma_{\frac{1}{2}, \frac{1}{2}})\chi^\sigma_{\frac{1}{2}, -\frac{1}{2}} \nonumber \\ 
&-(\sqrt{\frac{1}{3}}\chi^\sigma_{10}\chi^\sigma_{\frac{1}{2}, -\frac{1}{2}}-\sqrt{\frac{2}{3}}\chi^\sigma_{1, -1}\chi^\sigma_{\frac{1}{2}, \frac{1}{2}})\chi^\sigma_{\frac{1}{2}, \frac{1}{2}}\big) \,, \\
\chi_{0,0}^{\sigma_{u_4}}(4) &= \frac{1}{\sqrt{2}}(\chi^\sigma_{00}\chi^\sigma_{\frac{1}{2}, \frac{1}{2}}\chi^\sigma_{\frac{1}{2}, -\frac{1}{2}}-\chi^\sigma_{00}\chi^\sigma_{\frac{1}{2}, -\frac{1}{2}}\chi^\sigma_{\frac{1}{2}, \frac{1}{2}}) \,,
\end{align}
for $(S,M_S)=(0,0)$, by 
\begin{align}
\label{SWF1}
\chi_{1,1}^{\sigma_{w_1}}(4) &= \chi^\sigma_{00}\chi^\sigma_{11} \,, \\ 
\chi_{1,1}^{\sigma_{w_2}}(4) &= \chi^\sigma_{11}\chi^\sigma_{00} \,, \\
\chi_{1,1}^{\sigma_{w_3}}(4) &= \frac{1}{\sqrt{2}} (\chi^\sigma_{11} \chi^\sigma_{10}-\chi^\sigma_{10} \chi^\sigma_{11}) \,, \\
\chi_{1,1}^{\sigma_{w_4}}(4) &= \sqrt{\frac{3}{4}}\chi^\sigma_{11}\chi^\sigma_{\frac{1}{2}, \frac{1}{2}}\chi^\sigma_{\frac{1}{2}, -\frac{1}{2}}-\sqrt{\frac{1}{12}}\chi^\sigma_{11}\chi^\sigma_{\frac{1}{2}, -\frac{1}{2}}\chi^\sigma_{\frac{1}{2}, \frac{1}{2}} \nonumber \\ 
&-\sqrt{\frac{1}{6}}\chi^\sigma_{10}\chi^\sigma_{\frac{1}{2}, \frac{1}{2}}\chi^\sigma_{\frac{1}{2}, \frac{1}{2}} \,, \\
\chi_{1,1}^{\sigma_{w_5}}(4) &= (\sqrt{\frac{2}{3}}\chi^\sigma_{11}\chi^\sigma_{\frac{1}{2}, -\frac{1}{2}}-\sqrt{\frac{1}{3}}\chi^\sigma_{10}\chi^\sigma_{\frac{1}{2}, \frac{1}{2}})\chi^\sigma_{\frac{1}{2}, \frac{1}{2}} \,, \\
\chi_{1,1}^{\sigma_{w_6}}(4) &= \chi^\sigma_{00}\chi^\sigma_{\frac{1}{2}, \frac{1}{2}}\chi^\sigma_{\frac{1}{2}, \frac{1}{2}} \,,
\end{align}
for $(S,M_S)=(1,1)$, and by 
\begin{align}
\label{SWF2}
\chi_{2,2}^{\sigma_{1}}(4) &= \chi^\sigma_{11}\chi^\sigma_{11} \,,
\end{align}
for $(S,M_S)=(2,2)$. The superscripts $u_1,\ldots,u_4$ and $w_1,\ldots,w_6$ label the spin wave functions associated with each tetraquark configuration, as summarized in Table~\ref{SpinIndex}. These expressions are obtained by coupling the spins of the relevant subclusters using standard SU(2) algebra, with the basic two-body spin states defined as
\begin{align}
\label{Spin}
&\chi^\sigma_{00} = \frac{1}{\sqrt{2}}(\chi^\sigma_{\frac{1}{2}, \frac{1}{2}} \chi^\sigma_{\frac{1}{2}, -\frac{1}{2}}-\chi^\sigma_{\frac{1}{2}, -\frac{1}{2}} \chi^\sigma_{\frac{1}{2}, \frac{1}{2}}) \,, \\
&\chi^\sigma_{11} = \chi^\sigma_{\frac{1}{2}, \frac{1}{2}} \chi^\sigma_{\frac{1}{2}, \frac{1}{2}} \,, \\
&\chi^\sigma_{1,-1} = \chi^\sigma_{\frac{1}{2}, -\frac{1}{2}} \chi^\sigma_{\frac{1}{2}, -\frac{1}{2}} \,, \\
&\chi^\sigma_{10} = \frac{1}{\sqrt{2}}(\chi^\sigma_{\frac{1}{2}, \frac{1}{2}} \chi^\sigma_{\frac{1}{2}, -\frac{1}{2}}+\chi^\sigma_{\frac{1}{2}, -\frac{1}{2}} \chi^\sigma_{\frac{1}{2}, \frac{1}{2}}) \,.
\end{align}

\begin{table}[!t]
\caption{\label{SpinIndex} The values of the superscripts $u_1,\ldots,u_4$ and $w_1,\ldots,w_6$ that determine the spin wave function for each configuration of the fully strange tetraquark systems.}
\begin{ruledtabular}
\begin{tabular}{lcccccc}
& Di-meson & Diquark-antidiquark & $K_1$ & $K_2$ & $K_3$ & $K_4$ \\
\hline
$u_1$ & 1   & 3 & & & & \\
$u_2$ & 2  & 4 & & & & \\
$u_3$ &   &  & 5 & 7 & 9 & 11 \\
$u_4$ &   &  & 6 & 8 & 10 & 12 \\[2ex]
$w_1$  & 1   & 4 & & & & \\
$w_2$ & 2  & 5 & & & & \\
$w_3$ & 3  & 6 & & & & \\
$w_4$ &   &  & 7 & 10 & 13 & 16 \\
$w_5$ &   &  & 8 & 11 & 14 & 17 \\
$w_6$ &   &  & 9 & 12 & 15 & 18 \\
\end{tabular}
\end{ruledtabular}
\end{table}

For the five-body system, the spin wave functions of the baryon-meson configuration of Fig.~\ref{QQqq}(g) are given by
\begin{align}
\label{Spin1}
\chi_{\frac12,\frac12}^{\sigma 1}(5) &= \sqrt{\frac{1}{6}} \chi_{\frac32,-\frac12}^{\sigma}(3) \chi_{11}^{\sigma}
-\sqrt{\frac{1}{3}} \chi_{\frac32,\frac12}^{\sigma}(3) \chi_{10}^{\sigma} \nonumber \\
&
+\sqrt{\frac{1}{2}} \chi_{\frac32,\frac32}^{\sigma}(3) \chi_{1-1}^{\sigma} \,,  \\
\chi_{\frac12,\frac12}^{\sigma 2}(5) &= \sqrt{\frac{1}{3}} \chi_{\frac12,\frac12}^{\sigma+}(3) \chi_{10}^{\sigma} -\sqrt{\frac{2}{3}} \chi_{\frac12,-\frac12}^{\sigma+}(3) \chi_{11}^{\sigma} \,, \\
\chi_{\frac12,\frac12}^{\sigma 3}(5) &= \sqrt{\frac{1}{3}} \chi_{\frac12,\frac12}^{\sigma-}(3) \chi_{10}^{\sigma} - \sqrt{\frac{2}{3}} \chi_{\frac12,-\frac12}^{\sigma-}(3) \chi_{11}^{\sigma} \,, \\
\chi_{\frac12,\frac12}^{\sigma 4}(5) &= \chi_{\frac12,\frac12}^{\sigma+}(3) \chi_{00}^{\sigma} \,, \\
\chi_{\frac12,\frac12}^{\sigma 5}(5) &= \chi_{\frac12,\frac12}^{\sigma-}(3) \chi_{00}^{\sigma} \,, 
\end{align}
for $(S,M_S)=(1/2, 1/2)$, and by
\begin{align}
\label{Spin2}
\chi_{\frac32,\frac32}^{\sigma 1}(5) &= \sqrt{\frac{3}{5}}
\chi_{\frac32,\frac32}^{\sigma}(3) \chi_{10}^{\sigma} -\sqrt{\frac{2}{5}} \chi_{\frac32,\frac12}^{\sigma}(3) \chi_{11}^{\sigma} \,, \\
\chi_{\frac32,\frac32}^{\sigma 2}(5) &= \chi_{\frac32,\frac32}^{\sigma}(3) \chi_{00}^{\sigma} \,, \\
\chi_{\frac32,\frac32}^{\sigma 3}(5) &= \chi_{\frac12,\frac12}^{\sigma+}(3) \chi_{11}^{\sigma} \,, \\
\chi_{\frac32,\frac32}^{\sigma 4}(5) &= \chi_{\frac12,\frac12}^{\sigma-}(3) \chi_{11}^{\sigma} \,, 
\end{align}
for $(S,M_S)=(3/2, 3/2)$, and by
\begin{align}
\label{Spin3}
\chi_{\frac52,\frac52}^{\sigma 1}(5) &= \chi_{\frac32,\frac32}^{\sigma}(3) \chi_{11}^{\sigma} \,, 
\end{align}
for $(S,M_S)=(5/2, 5/2)$. 

The corresponding spin wave functions for the diquark-diquark-antiquark configuration [Fig.~\ref{QQqq}(h)] are given by
\begin{align}
\label{Spin4}
\chi_{\frac12,\frac12}^{\sigma 6}(5) &= \chi_{00}^{\sigma} \chi_{00}^{\sigma} \chi_{\frac12,\frac12}^{\sigma} \,, \\
\chi_{\frac12,\frac12}^{\sigma 7}(5) &= \sqrt{\frac{2}{3}}\chi_{00}^{\sigma} \chi_{11}^{\sigma} \chi_{\frac12,-\frac12}^{\sigma}-\sqrt{\frac{1}{3}}\chi_{00}^{\sigma} \chi_{10}^{\sigma} \chi_{\frac12,\frac12}^{\sigma} \,, \\
\chi_{\frac12,\frac12}^{\sigma 8}(5) &= \sqrt{\frac{1}{3}}(\chi_{11}^{\sigma} \chi_{1-1}^{\sigma}-\chi_{10}^{\sigma} \chi_{10}^{\sigma}+\chi_{1-1}^{\sigma} \chi_{11}^{\sigma})\chi_{\frac12,\frac12}^{\sigma} \,, \\
\chi_{\frac12,\frac12}^{\sigma 9}(5) &= \sqrt{\frac{1}{3}}(\chi_{11}^{\sigma} \chi_{10}^{\sigma}-\chi_{10}^{\sigma} \chi_{11}^{\sigma})\chi_{\frac12,-\frac12}^{\sigma}- \\ \nonumber
&
\sqrt{\frac{1}{6}}(\chi_{11}^{\sigma} \chi_{1-1}^{\sigma}-\chi_{1-1}^{\sigma} \chi_{11}^{\sigma})\chi_{\frac12,\frac12}^{\sigma} \,, 
\end{align}
for $(S,M_S)=(1/2, 1/2)$, and by
\begin{align}
\label{Spin5}
\chi_{\frac32,\frac32}^{\sigma 5}(5) &= \chi_{00}^{\sigma} \chi_{11}^{\sigma} \chi_{\frac12,\frac12}^{\sigma} \,, \\
\chi_{\frac32,\frac32}^{\sigma 6}(5) &= \sqrt{\frac{1}{2}}(\chi_{11}^{\sigma} \chi_{10}^{\sigma}-\chi_{10}^{\sigma} \chi_{11}^{\sigma})\chi_{\frac12,\frac12}^{\sigma} \,, \\
\chi_{\frac32,\frac32}^{\sigma 7}(5) &= \sqrt{\frac{4}{5}}\chi_{11}^{\sigma} \chi_{11}^{\sigma}\chi_{\frac12,-\frac12}^{\sigma}- \nonumber \\
&
\sqrt{\frac{1}{10}}(\chi_{11}^{\sigma} \chi_{10}^{\sigma}+\chi_{10}^{\sigma} \chi_{11}^{\sigma})\chi_{\frac12,\frac12}^{\sigma} \,, 
\end{align}
for $(S,M_S)=(3/2, 3/2)$, and by
\begin{align}
\label{Spin6}
\chi_{\frac52,\frac52}^{\sigma 2}(5) &= \chi_{11}^{\sigma} \chi_{11}^{\sigma} \chi_{\frac12,\frac12}^{\sigma} \,, 
\end{align}
for $(S,M_S)=(5/2, 5/2)$. These spin bases have been extensively used and validated in our previous studies of multiquark systems~\cite{Yang:2015bmv, Yang:2018oqd, gy:2020dcp}.

The flavor wave functions of the fully strange tetra- and pentaquark systems are trivial and are written as
\begin{align}
\chi^{f_1} = \bar{s}s\bar{s}s \,, \quad \chi^{f_2} = sss\bar{s}s \,.
\end{align}

The four- and five-body Schr\"odinger equations are solved using the Rayleigh-Ritz variational principle. Within the complex-scaled framework, the spatial wave function of a tetraquark system is expanded as
\begin{equation}
\label{eq:WFexp}
\psi^T_{LM_L}= \left[ \left[ \phi_{n_1l_1}(\vec{\rho}e^{i\theta}\,) \phi_{n_2l_2}(\vec{\lambda}e^{i\theta}\,)\right]_{l} \phi_{n_3l_3}(\vec{R}e^{i\theta}\,) \right]_{L M_L} \,,
\end{equation}
where the Jacobi coordinates are defined according to the specific configuration under consideration. That is to say, we have
\begin{align}
\vec{\rho} &= \vec{x}_1-\vec{x}_{2(4)} \,, \\
\vec{\lambda} &= \vec{x}_3 - \vec{x}_{4(2)} \,, \\
\vec{R} &= \frac{m_1 \vec{x}_1 + m_{2(4)} \vec{x}_{2(4)}}{m_1+m_{2(4)}}- \frac{m_3 \vec{x}_3 + m_{4(2)} \vec{x}_{4(2)}}{m_3+m_{4(2)}} \,,
\end{align}
for the meson-meson configuration of Fig.~\ref{QQqq}(a), and
\begin{align}
\vec{\rho} &= \vec{x}_1-\vec{x}_3 \,, \\
\vec{\lambda} &= \vec{x}_2 - \vec{x}_4 \,, \\
\vec{R} &= \frac{m_1 \vec{x}_1 + m_3 \vec{x}_3}{m_1+m_3}- \frac{m_2 \vec{x}_2 + m_4 \vec{x}_4}{m_2+m_4} \,,
\end{align}
for the diquark-antidiquark arrangement of Fig.~\ref{QQqq}(b). The other four K-type configurations presented in Fig.~\ref{QQqq}(c)--(f) are ($i, j, k, l$ take values according to the panels (c) to (f) of Fig.~\ref{QQqq}):
\begin{align}
\vec{\rho} &= \vec{x}_i-\vec{x}_j \,, \\
\vec{\lambda} &= \vec{x}_k- \frac{m_i \vec{x}_i + m_j \vec{x}_j}{m_i+m_j} \,, \\
\vec{R} &= \vec{x}_l- \frac{m_i \vec{x}_i + m_j \vec{x}_j+m_k \vec{x}_k}{m_i+m_j+m_k} \,.
\end{align}

Similarly, the spatial wave function of the pentaquark system is written as
\begin{align}
\label{eq:WFexpP}
&
\psi^P_{LM_L}=[ [ [ \phi_{n_1l_1}({\vec{\rho}} e^{i\theta})\phi_{n_2l_2}({\vec{\lambda}} e^{i\theta})]_{l} \phi_{n_3l_3}({\vec{r}} e^{i\theta}) ]_{l^{\prime}} \nonumber\\
&
\hspace*{1.60cm}  \phi_{n_4l_4}({\vec{R}} e^{i\theta}) ]_{LM_L} \,,
\end{align}
with the corresponding Jacobi coordinates defined for the baryon-meson configuration of Fig.~\ref{QQqq}(g) given by
\begin{align}
\vec{\rho} &= \vec{x}_1-\vec{x}_2 \,, \\
\vec{\lambda} &= \vec{x}_3 - \frac{{m_1\vec{x}}_1+{m_2\vec{x}}_2}{m_1+m_2} \,,  \\
\vec{r} &= \vec{x}_4-\vec{x}_5 \,, \\
\vec{R} &= \frac{m_1\vec{x}_1+m_2\vec{x}_2 + m_3\vec{x}_3}{m_1+m_2+m_3}-\frac{m_4\vec{x}_4+m_5\vec{x}_5}{m_4+m_5} \,,
\end{align}
and for the diquark-diquark-antiquark one of Fig.~\ref{QQqq}(h) provided by
\begin{align}
\vec{\rho} &= \vec{x}_1-\vec{x}_2 \,, \\
\vec{\lambda} &= \vec{x}_3-\vec{x}_5 \,, \\
\vec{r} &= \frac{m_1\vec{x}_1+m_2\vec{x}_2}{m_1+m_2}-\frac{m_3\vec{x}_3+m_5\vec{x}_5}{m_3+m_5} \,,\\
\vec{R} &= \vec{x}_4-\frac{m_1\vec{x}_1+m_2\vec{x}_2 + m_3\vec{x}_3+m_5\vec{x}_5}{m_1+m_2+m_3+m_5}\,.
\end{align}
Note that the center-of-mass kinetic energy $T_{\text{CM}}$ is exactly removed by working in relative Jacobi coordinates. 

The Gaussian Expansion Method (GEM)~\cite{Hiyama:2003cu} is employed to expand the spatial wave functions, expressed in Eq.~\eqref{eq:WFexp} and Eq.~\eqref{eq:WFexpP}, with Gaussian ranges chosen in geometric progression to ensure numerical convergence. Each relative-motion wave function is expanded in a Gaussian basis, 
\begin{align}
\phi_{nlm}(\vec{r}e^{i\theta}\,) = \sqrt{1/4\pi} \, N_{nl} \, (re^{i\theta})^{l} \, e^{-\nu_{n} (re^{i\theta})^2} \,.
\end{align}
Further details of the method can be found in Ref.~\cite{Yang:2015bmv}. For S-wave states, the orbital wave functions take the simplified form
\begin{align}
\phi_{n00}(\vec{r}e^{i\theta}\,) = \sqrt{1/4\pi} \, N_{n0} \, e^{-\nu_{n} (re^{i\theta})^2} \,.
\end{align}
The size parameters $\nu_n$ are defined as~\cite{Hiyama:2003cu}:
\begin{align}
\nu_n = r_n^{-2} \,,\quad r_n=r_1 a^{n-1} \,.
\end{align}
The numerical values of $(r_1,a)$ used for tetraquark and pentaquark systems are specified to optimize convergence. Particularly, in the spatial wave function of tetraquark, values of $(r_1, a)$ are $(0.01,2.0)$ for the $\vec{\rho}$ and $\vec{\lambda}$ relative motions, whereas for the relative motion $\vec{R}$ they are $(0.1,4.0)$. In the pentaquark case, they are $(0.1,2.0)$ for the first three relative motions, $\vec{\rho}$, $\vec{\lambda}$ and $\vec{r}$. The last relative motion, $\vec{R}$, takes the same values of $(0.1,4.0)$ as those in the 4-body system.

The complete tetraquark ($T$) and pentaquark ($P$) wave functions are finally constructed as
\begin{align}
\label{TPs}
 \Psi^{T(P)}_{J M_J} &= \sum_{i, j} c_{ij} \Psi^{T(P)}_{J M_J, i, j} \nonumber \\
 &=\sum_{i, j} c_{ij} {\cal A}^{T(P)} \left[ \left[ \psi^{T(P)}_{L M_L} \chi^{\sigma_i}_{S M_S} \right]_{J M_J} \chi^{f} \chi^{c}_j \right] \,,
\end{align}
where ${\cal{A}}^{T(P)}$ is the antisymmetry operator of a multi-quark system, which take into account the fact of involving two identical strange (anti-)quarks. Particularly, the definition of tetraquark system according to Fig.~\ref{QQqq} is 
\begin{equation}
\label{Antisym}
{\cal{A}}^T = 1-(13)-(24)+(13)(24) \,.
\end{equation}
Meanwhile, in the pentaquark system, the baryon-meson structure of Fig.~\ref{QQqq}(g) and the diquark-diquark-antiquark arrangement of Fig.~\ref{QQqq}(h) have the following antisymmetry operators, respectively.
\begin{align}
&
{\cal{A}}^P_1 = [1-(13)-(23)][1-(15)-(25)-(35)] \,, \label{EE1} \\
&
{\cal{A}}^P_2 = 1-(13)-(15)-(23)-(25)+(13)(25) \,. \label{EE2}
\end{align}
Note here that, since the total wave function is constructed from multiple subcluster configurations, proper antisymmetrization is essential to satisfy the Pauli principle.

The nature of fully strange tetra- and penta-quark systems are further investigated by quantitative analyses of inter-quark distances,
\begin{equation}
\label{quarkdistance}
{r_{q\bar{q}}} = Re(\sqrt{\langle \Psi^{T(P)}_{J M_J} \vert  (r_{q\bar{q}} e^{i\theta})^2 \vert \Psi^{T(P)}_{J M_J} \rangle}) \,,
\end{equation}
magnetic moments,
\begin{equation}
\label{MM}
{\mu_m} = Re(\langle \Psi^{T(P)}_{J M_J} \vert  \sum_{i=1}^{n} \frac{\hat Q_i}{2m_i} \hat \sigma^z_i \vert \Psi^{T(P)}_{J M_J} \rangle) \,,
\end{equation}
and a qualitative survey of dominant wavefunction components,
\begin{equation}
\label{DCC}
C_p = Re(\sum_{i,j} \langle c^l_{ij} \Psi^{T(P)}_{J M_J, i, j} \vert c^r_{ij} \Psi^{T(P)}_{J M_J, i, j} \rangle) \,.
\end{equation}
In Eq.~\eqref{MM}, $\hat Q_i$ and $\hat\sigma^z_i$ denote the electric charge and spin operators of the $i$th quark, respectively. The coefficients $c^l_{ij}$ and $c^r_{ij}$ correspond to the left and right generalized eigenvectors of the complex-scaled Hamiltonian.

Since different multiquark channels are not mutually orthogonal within the present framework, two complementary procedures are employed to evaluate the contributions of individual configurations. In the first approach, only the diagonal elements of Eq.~(\ref{DCC}) are considered. In the second, off-diagonal elements are incorporated by summing all elements in a given row and attributing them to the corresponding diagonal contribution, thereby accounting for interference effects between configurations. A comparison of these two methods allows us to reliably identify dominant components and assess the impact of channel mixing in both bound and resonant states.


\section{Results}
\label{sec:results}

This section presents the results and discussion for fully strange tetra- and penta-quark systems. As a first step, the spectra are analyzed on the real energy axis by setting the complex rotation angle to $\theta=0^\circ$. In a fully coupled-channel calculation, however, bound and resonant states are generally embedded in the continuum. To clearly distinguish bound, resonant, and scattering states, we employ the Complex Scaling Method (CSM), in which these different classes of states are separated in the complex energy plane once a nonzero rotation angle $\theta$ is introduced. To reliably extract well-defined eigenvalues and eigenvectors for multiquark systems including all $S$-wave configurations, the rotation angle is systematically varied from $0^\circ$ to $14^\circ$, serving as an effective auxiliary parameter within the CSM framework.

The calculated low-lying mass spectra of fully strange tetra- and penta-quarks are summarized in Tables~\ref{GresultCC1}–\ref{GresultCCT}. We first present the real-axis results for each allowed $J^P$ quantum number. The corresponding lowest theoretical masses are listed in Tables~\ref{GresultCC1},~\ref{GresultCC2},~\ref{GresultCC3},~\ref{GresultCC4},~\ref{GresultCC5} and \ref{GresultCC6}. In these tables, the considered configurations -- meson-meson, diquark-antidiquark, and $K$-type, meson-baryon and diquark-diquark-quark -- are given in the first column. When applicable, the corresponding noninteracting meson-meson or baryon-meson thresholds, taken from experimental data, are shown in parentheses. Each channel is labeled by an index in the second column, which specifies a particular combination of spin ($\chi_J^{\sigma_i}$) and color ($\chi_j^c$) wave functions, explicitly listed in the third column. The fourth column reports the lowest mass obtained for each individual channel, while the final column gives the result of the coupled-channel calculation for the corresponding configuration. The last row of each table presents the lowest mass obtained from the full coupled-channel calculation on the real energy axis and, when applicable, the associated binding energy of the bound state.

\begin{figure}[!t]
\includegraphics[clip, trim={3.0cm 1.7cm 3.0cm 0.8cm}, width=0.45\textwidth]{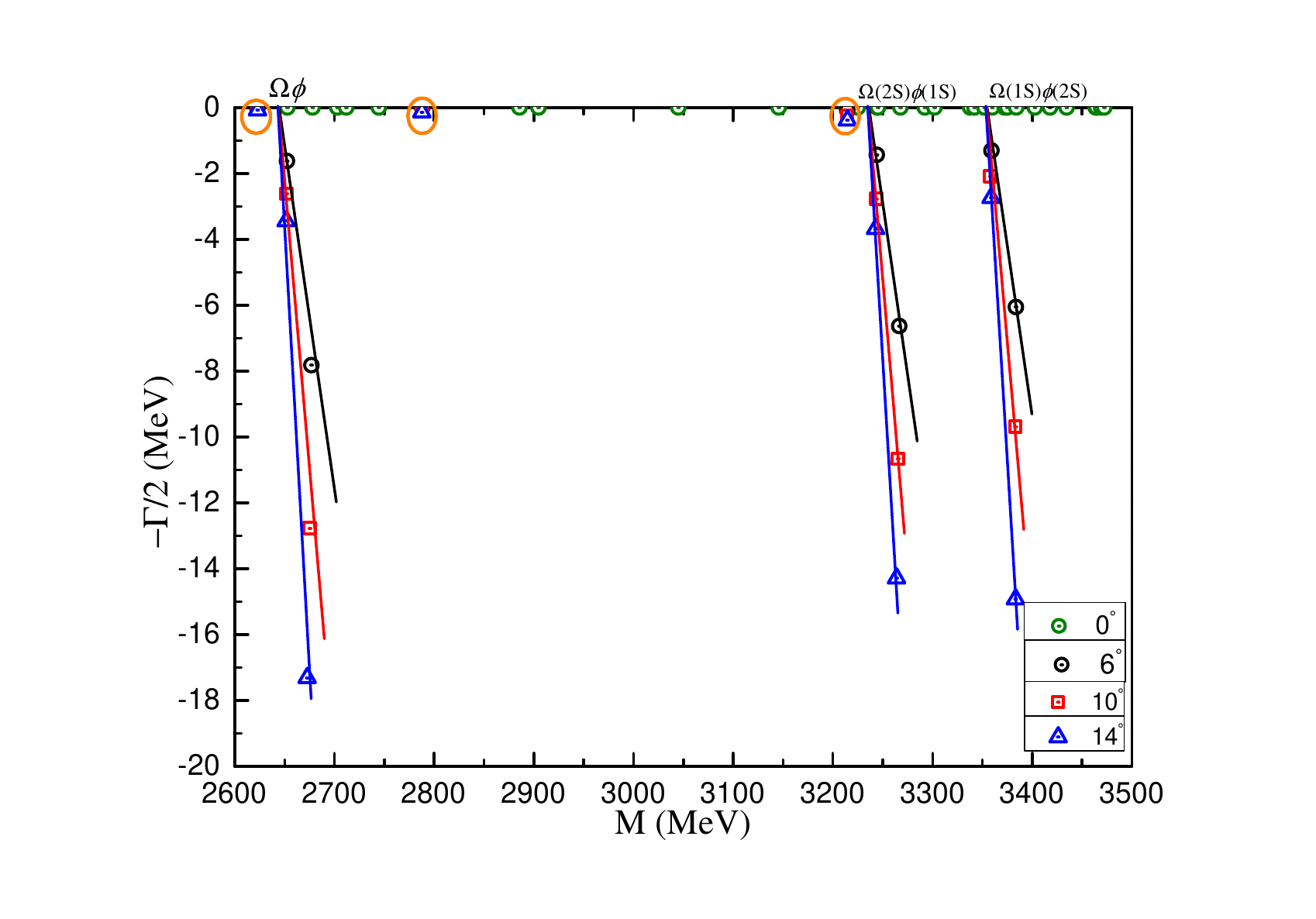}
\caption{\label{PP4} The fully coupled-channels calculation of $ssss\bar{s}$ pentaquark system with $J^P=\frac{1}{2}^-$ quantum numbers.}
\end{figure}

Subsequently, full coupled-channel calculations for each allowed $J^P$ quantum number are carried out within the CSM. The resulting distributions of complex eigenenergies are shown in Figures~\ref{PP1}--\ref{PP6}, where the identified bound and resonant states are highlighted by circles. The properties of the extracted exotic states are summarized in Tables~\ref{GresultR1},~\ref{GresultR2},~\ref{GresultR3},~\ref{GresultR4},~\ref{GresultR5} and~\ref{GresultR6}, including their root-mean-square radii, magnetic moments, and dominant configuration components. Finally, the most significant results of the present study are collected in Table~\ref{GresultCCT}.

\subsection{Fully strange tetraquarks}


\begin{table}[!t]
\caption{\label{GresultCC1} Lowest-lying $ss\bar{s}\bar{s}$ tetraquark states with $J^P=0^+$ calculated within the real range formulation of the chiral quark model.
The allowed meson-meson, diquark-antidiquark and K-type configurations are listed in the first column, the experimental value of the non-interacting meson-meson threshold is labeled in parentheses. Each channel is assigned an index in the 2nd column, it reflects a particular combination of spin ($\chi_J^{\sigma_i}$) and color ($\chi_j^c$) wave functions that are shown explicitly in the 3rd column. The theoretical mass obtained in each channel is shown in the 4th column and the coupled result for each kind of configuration  is presented in the 5th column.
When a complete coupled-channels calculation is performed, last row of the table indicates the calculated lowest-lying mass and binding energy (unit: MeV).}
\begin{ruledtabular}
\begin{tabular}{lcccc}
~~Channel   & Index & $\chi_J^{\sigma_i}$;~$\chi_j^c$ & $M$ & Mixed~~ \\
        &   &$[i; ~j]$ &  \\[2ex]
$(\eta' \eta')^1 (1916)$  & 1  & [1;~1]  & $1656$ & \\
$(\phi \phi)^1 (2040)$  & 2  & [2;~1]   & $2022$ & $1656$  \\[2ex]
$(\eta' \eta')^8$          & 3  & [1;~2]  & $2360$ & \\
$(\phi \phi)^8$       & 4  & [2;~2]   & $2339$ & $2135$  \\[2ex]
$(ss)(\bar{s}\bar{s})$      & 5     & [3;~4]  & $2333$ & \\
$(ss)^*(\bar{s}\bar{s})^*$  & 6  & [4;~3]   & $2225$ & $2114$ \\[2ex]
$K_1$  & 7  & [5;~5]   & $2346$ & \\
            & 8  & [5;~6]   & $2134$ & \\
            & 9  & [6;~5]   & $2366$ & \\
            & 10  & [6;~6]   & $2278$ & $2121$ \\[2ex]
$K_2$  & 11  & [7;~7]   & $2134$ & \\
             & 12  & [7;~8]   & $2346$ & \\
             & 13  & [8;~7]   & $2278$ & \\
             & 14  & [8;~8]   & $2366$ & $2121$ \\[2ex]
$K_3$  & 15  & [9;~10]   & $2226$ & \\
             & 16  & [10;~9]   & $2329$ & $2106$ \\[2ex]
$K_4$  & 17  & [11;~12]   & $2226$ & \\
             & 18  & [12;~11]   & $2329$ & $2106$ \\[2ex]
\multicolumn{4}{c}{Complete coupled-channels:} & $1645$ \\
\multicolumn{4}{c}{} & $E_B=-11$
\end{tabular}
\end{ruledtabular}
\end{table}

\begin{figure}[!t]
\includegraphics[clip, trim={3.0cm 1.7cm 3.0cm 0.8cm}, width=0.45\textwidth]{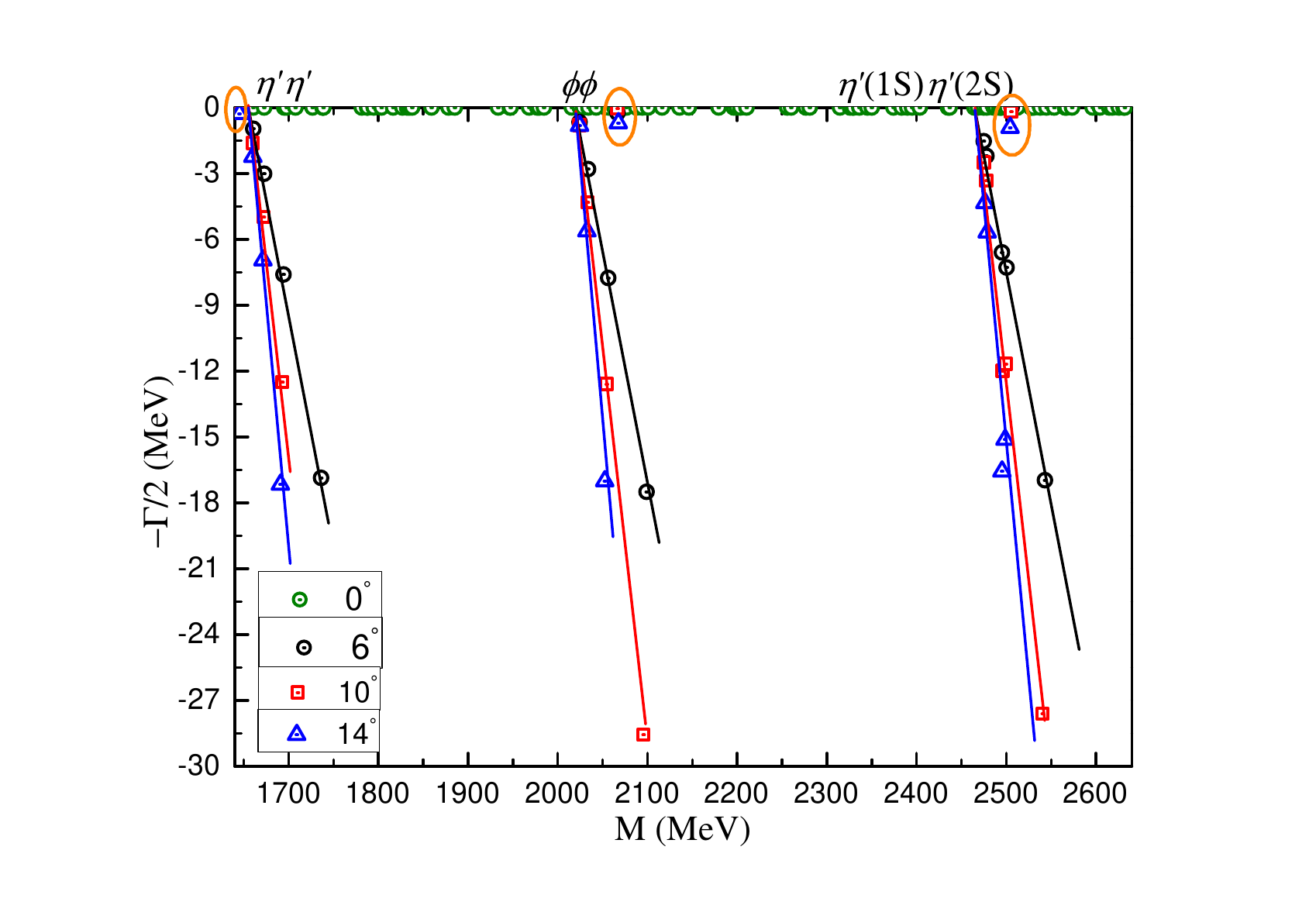}
\caption{\label{PP1} The fully coupled-channels calculation of $ss\bar{s}\bar{s}$ tetraquark system with $J^P=0^+$ quantum numbers.}
\end{figure}

\begin{table}[!t]
\caption{\label{GresultR1} Compositeness of exotic states obtained in a complete coupled-channels calculation in the $0^+$ state of $ss\bar{s}\bar{s}$ tetraquark. Particularly, the first column is exotic poles labeled by $M-i\Gamma$, unit in MeV; the second one is the magnetic moment of state and, if possible, the binding energy of bound state, unit in $\mu_N$ and MeV, respectively; the distance between any two quarks or quark-antiquark, unit in fm; and the component of exotic state ($S$: dimeson structure in color-singlet channel; $H$: dimeson structure in hidden-color channel; $Di$: diquark-antiquark configuration; $K$: K-type configuration). Herein, two sets of results on a exotic state component are listed. Particularly, set I is results on components, that only diagonal elements are employed, and results, that both diagonal and off-diagonal elements are considered, are listed in set II.}
\begin{ruledtabular}
\begin{tabular}{rccc}
Exotic state       & \multicolumn{3}{c}{Structure} \\[2ex]
$1645-i0$   & \multicolumn{3}{c}{$\mu=0$, $E_B=-11$} \\
  & \multicolumn{3}{c}{$r_{s \bar{s}}:1.13$;\,\,\,\,\,$r_{\bar{s}\bar{s}}:1.21$;\,\,\,\,\,$r_{ss}:1.21$} \\
$Set$ I: & \multicolumn{3}{c}{$S$: 9.1\%;\, $H$: 1.4\%;\, $Di$: 37.5\%;\, $K$: 52\%}\\
$Set$ II: & \multicolumn{3}{c}{$S$: 10.5\%;\, $H$: 1.5\%;\, $Di$: 33.5\%;\, $K$: 54.5\%}\\[2ex]
$2067-i0.1$   & \multicolumn{3}{c}{$\mu=0$} \\
  & \multicolumn{3}{c}{$r_{s \bar{s}}:0.73$;\,\,\,\,\,$r_{\bar{s}\bar{s}}:1.06$;\,\,\,\,\,$r_{ss}:1.21$} \\
$Set$ I: & \multicolumn{3}{c}{$S$: 0.8\%;\, $H$: 0.4\%;\, $Di$: 62.4\%;\, $K$: 36.4\%}\\
$Set$ II: & \multicolumn{3}{c}{$S$: 2.1\%;\, $H$: 1.6\%;\, $Di$: 62.5\%;\, $K$: 33.8\%}\\[2ex]
$2506-i0.4$   & \multicolumn{3}{c}{$\mu=0$} \\
  & \multicolumn{3}{c}{$r_{s \bar{s}}:1.19$;\,\,\,\,\,$r_{\bar{s}\bar{s}}:1.43$;\,\,\,\,\,$r_{ss}:1.43$} \\
$Set$ I: & \multicolumn{3}{c}{$S$: 0\%;\, $H$: 0\%;\, $Di$: 0.1\%;\, $K$: 99.9\%}\\
$Set$ II: & \multicolumn{3}{c}{$S$: 0\%;\, $H$: 0\%;\, $Di$: 0.1\%;\, $K$: 99.9\%}\\
\end{tabular}
\end{ruledtabular}
\end{table}

{\bf The $\bm{J^P=0^+}$ sector:}
A total of 18 channels are considered in the $J^P=0^+$ sector, as listed in Table~\ref{GresultCC1}. We begin with the two meson-meson configurations, $\eta' \eta'$ and $\phi\phi$, each examined in both color-singlet and hidden-color channels. In the color-singlet case, neither configuration is bound, and the lowest masses coincide with their corresponding theoretical thresholds, namely $1.66$ GeV for $\eta' \eta'$ and $2.02$ GeV for $\phi\phi$. The hidden-color counterparts lie significantly higher, at around $2.35$ GeV. For the remaining exotic configurations: two diquark-antidiquark channels and twelve $K$-type channels, the calculated masses fall in the range $2.13-2.37$ GeV. Consequently, no bound state is found in any single-channel calculation. Moreover, the $K_1$ and $K_2$ configurations are found to be degenerate in mass, a feature that also appears in the $K_3$ and $K_4$ cases and persists in the $J^P=1^+$ and $2^+$ sectors of fully strange tetraquarks.

We next perform coupled-channel calculations within each of the seven configuration classes, namely the meson-meson structures in both color-singlet and color-octet channels, the diquark-antidiquark arrangement, and the four $K$-type configurations. In all these cases, the lowest mass remains at $1.66$ GeV, corresponding to the $\eta'\eta'$ threshold, while the remaining coupled states with exotic color structures cluster around $2.11$ GeV. However, when all $18$ channels are fully coupled, a shallow bound state emerges on the real energy axis at $\theta=0^\circ$, with a mass of $1645$ MeV and a binding energy of $E_B=-11$ MeV.

A more refined analysis is carried out using the Complex Scaling Method. The resulting complex energy spectrum is displayed in Fig.~\ref{PP1}. Within the mass range $1.64-2.65$ GeV, three continuum trajectories corresponding to the $\eta'\eta'$, $\phi\phi$, and $\eta'(1S)\eta'(2S)$ scattering states are clearly identified. In addition, three isolated poles that remain stable under variations of the rotation angle $\theta$ are observed. The first pole, located at $1645$ MeV with $E_B=-11$ MeV, corresponds to the bound state identified in the real-axis analysis. The remaining two poles represent extremely narrow resonances with complex energies $M-i\Gamma = 2067 - i\,0.1$ MeV and $2506 - i\,0.4$ MeV, respectively.

The properties of these three exotic states are summarized in Table~\ref{GresultR1}. All of them have vanishing magnetic moments, a feature that is also observed in the $J^P=1^+$ and $2^+$ fully strange tetraquark sectors. The root-mean-square radii of the bound state and the lower resonance are both approximately $1.1$ fm, whereas the higher resonance exhibits a larger spatial extent of about $1.4$ fm. An analysis of the wave-function composition shows that the dominant components arise from either diquark-antidiquark or $K$-type configurations. In particular, strong mixing between these two structures is found for the bound state and the lower resonance, while the higher resonance is almost entirely dominated by a single $K$-type component.

The shallow bound state with $E_B=-11$ MeV is expected to manifest most clearly in the $\eta'\eta'$ channel. After correcting the mass using the experimental $\eta'\eta'$ threshold, the modified mass is estimated to be 1905 MeV\footnote{The modified mass $M'$ is obtained by adding the theoretical binding energy $E_B$ to the experimental two-meson threshold $M_{\mathrm{exp}}$, i.e., $M'=M_{\mathrm{exp}}+E_B$.}. The two narrow resonances with exotic color configurations may also be accessible in future high-energy experiments through analyses of $\phi\phi$ or $\eta'\eta'$ two-body strong decay channels.


\begin{table}[!t]
\caption{\label{GresultCC2} Lowest-lying $ss\bar{s}\bar{s}$ tetraquark states with $J^P=1^+$ calculated within the real range formulation of the constituent quark model. Results are similarly organized as those in Table~\ref{GresultCC1} (unit: MeV).}
\begin{ruledtabular}
\begin{tabular}{lcccc}
~~Channel   & Index & $\chi_J^{\sigma_i}$;~$\chi_j^c$ & $M$ & Mixed~~ \\
        &   &$[i; ~j]$ &  \\[2ex]
$(\eta' \phi)^1 (1978)$   & 1  & [1;~1]  & $1839$ & \\
$(\phi \phi)^1 (2040)$    & 2  & [3;~1]   & $2022$ & $1839$  \\[2ex]
$(\eta' \phi)^8$             & 3  & [1;~2]   & $2118$ &  \\
$(\phi \phi)^8$               & 4  & [3;~2]   & $2285$ & $2118$ \\[2ex]
$(ss)^*(\bar{s}\bar{s})^*$  & 5  & [6;~3]   & $2218$ & $2218$  \\[2ex]
$K_1$      & 6  & [7;~5]   & $2215$ &  \\
                & 7 & [8;~5]    & $2190$ & \\
                & 8  & [9;~5]   & $2155$ & \\
                & 9   & [7;~6]  & $2211$ & \\
                & 10   & [8;~6]  & $2201$ & \\
                & 11   & [9;~6]  & $2092$ & $2014$ \\[2ex]
$K_2$      & 12  & [10;~7]   & $2211$ & \\
                 & 13  & [11;~7]   & $2201$ &  \\
                 & 14  & [12;~7]   & $2092$ & \\
                 & 15  & [10;~8]   & $2215$ & \\
                 & 16  & [11;~8]   & $2190$ & \\
                 & 17  & [12;~8]   & $2155$ & $2014$ \\[2ex]
$K_3$      & 18  & [13;~10]   & $2224$ & \\
                 & 19  & [14;~10]   & $2223$ & \\
                 & 20  & [15;~9]    & $2909$ & $2213$ \\[2ex]
$K_4$      & 21  & [16;~12]   & $2224$ & \\
                 & 22  & [17;~12]   & $2223$ & \\
                 & 23  & [18;~11]   & $2909$ & $2213$ \\[2ex]
\multicolumn{4}{c}{Complete coupled-channels:} & $1839$
\end{tabular}
\end{ruledtabular}
\end{table}

\begin{figure}[!t]
\includegraphics[clip, trim={3.0cm 1.7cm 3.0cm 0.8cm}, width=0.45\textwidth]{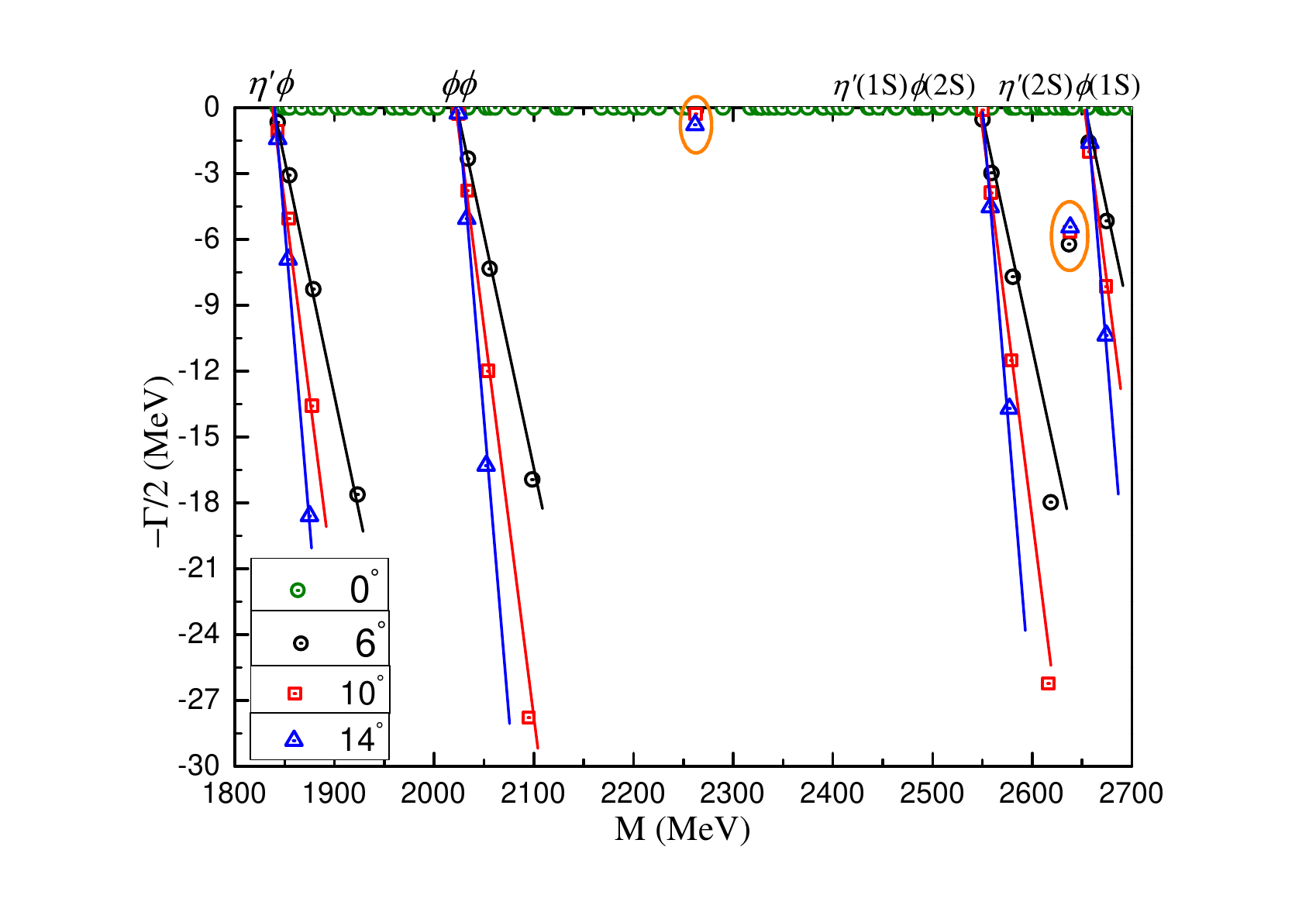}
\caption{\label{PP2} The fully coupled-channels calculation of $ss\bar{s}\bar{s}$ tetraquark system with $J^P=1^+$ quantum numbers.}
\end{figure}

\begin{table}[!t]
\caption{\label{GresultR2} Compositeness of resonance obtained in a complete coupled-channel calculation in the $1^+$ state of $ss\bar{s}\bar{s}$ tetraquark. Results are similarly organized as those in Table~\ref{GresultR1}.}
\begin{ruledtabular}
\begin{tabular}{rccc}
Resonance       & \multicolumn{3}{c}{Structure} \\[2ex]
$2262-i0.6$   & \multicolumn{3}{c}{$\mu=0$} \\
  & \multicolumn{3}{c}{$r_{s \bar{s}}:0.79$;\,\,\,\,\,$r_{\bar{s}\bar{s}}:0.73$;\,\,\,\,\,$r_{ss}:0.73$} \\
$Set$ I: & \multicolumn{3}{c}{$S$: 18.2\%;\, $H$: 81.5\%;\, $Di$: 0.1\%;\, $K$: 0.2\%}\\
$Set$ II: & \multicolumn{3}{c}{$S$: 18.2\%;\, $H$: 81.3\%;\, $Di$: 0.1\%;\, $K$: 0.4\%}\\[2ex]
$2638-i11.3$   & \multicolumn{3}{c}{$\mu=0$} \\
  & \multicolumn{3}{c}{$r_{s \bar{s}}:1.31$;\,\,\,\,\,$r_{\bar{s}\bar{s}}:1.50$;\,\,\,\,\,$r_{ss}:1.50$} \\
$Set$ I: & \multicolumn{3}{c}{$S$: 55.1\%;\, $H$: 11.7\%;\, $Di$: 30.1\%;\, $K$: 3.1\%}\\
$Set$ II: & \multicolumn{3}{c}{$S$: 40.7\%;\, $H$: 2.9\%;\, $Di$: 20.0\%;\, $K$: 36.4\%}\\
\end{tabular}
\end{ruledtabular}
\end{table}

{\bf The $\bm{J^P=1^+}$ sector:}
A total of 23 channels are investigated in the $J^P=1^+$ sector, as summarized in Table~\ref{GresultCC2}. The meson-meson configurations consist of the $\eta'\phi$ and $\phi\phi$ channels, each considered in both color-singlet and hidden-color representations. In the color-singlet case, the lowest masses coincide with the corresponding theoretical thresholds, namely $1839$ MeV for $\eta'\phi$ and $2022$ MeV for $\phi\phi$. Their hidden-color counterparts are found to be approximately $270$ MeV heavier. Only one diquark-antidiquark configuration, $(ss)^*(\bar{s}\bar{s})^*$, is included in this sector, yielding a mass of $2218$ MeV. The remaining $18$ channels arise from four $K$-type configurations and predominantly populate the energy region $2.09-2.22$ GeV, with the exception of two channels in the $K_3$ and $K_4$ structures, whose masses are both around $2.91$ GeV. As in the $J^P=0^+$ sector, degeneracies are observed between the $K_1$ and $K_2$ configurations, as well as between $K_3$ and $K_4$.

Partial and full coupled-channel calculations performed on the real energy axis at $\theta=0^\circ$ do not produce any bound state in this sector. The lowest mass remains at $1.83$ GeV, corresponding to the $\eta'\phi$ threshold, while the other exotic configurations lie in the range $2.01-2.21$ GeV. The Complex Scaling Method is then applied to the fully coupled-channel system. The resulting complex energy spectrum is displayed in Fig.~\ref{PP2}. Within the mass region $1.8-2.7$ GeV, the scattering continua associated with the $\eta'\phi$ and $\phi\phi$ ground states, as well as their first radial excitations, $\eta'(1S)\phi(2S)$ and $\eta'(2S)\phi(1S)$, are clearly identified. Among the scattering trajectories, two isolated poles that remain stable under variations of the rotation angle $\theta$ are observed. These correspond to two resonant states with complex energies $M-i\Gamma = 2262 - i\,0.6$ MeV and $2638 - i\,11.3$ MeV, respectively.

The properties of these two exotic states are summarized in Table~\ref{GresultR2}. Although both states have vanishing magnetic moments, their internal structures differ markedly. The lower resonance at $2.26$ GeV is a compact $ss\bar{s}\bar{s}$ tetraquark, with a root-mean-square radius smaller than $0.8$ fm. In contrast, the higher resonance at $2.64$ GeV has a much larger spatial extent of approximately $1.5$ fm. An analysis of the wave-function composition shows that the lower resonance is dominated by hidden-color components, which account for about 82\% of its structure. For the higher resonance, color-singlet channels contribute more than 40\%, accompanied by strong mixing among all tetraquark configurations, including singlet and hidden-color meson-meson, diquark-antidiquark, and $K$-type structures.

Based on its mass, compact size, and dominant hidden-color composition, the experimentally observed $X(2300)$ resonance~\cite{BESIII:2025wpp} can be naturally interpreted as a compact fully strange tetraquark with $J^P=1^+$. This interpretation is consistent with recent theoretical studies~\cite{Wan:2025xhf, Cao:2025dze, Liu:2026ljb}. Both resonant states identified here are expected to be accessible in future high-energy experiments, with the most favorable decay channels being $\phi\phi$ and $\eta'\phi$, respectively.


\begin{table}[!t]
\caption{\label{GresultCC3} Lowest-lying $ss\bar{s}\bar{s}$ tetraquark states with $J^P=2^+$ calculated within the real range formulation of the constituent quark model. Results are similarly organized as those in Table~\ref{GresultCC1} (unit: MeV).}
\begin{ruledtabular}
\begin{tabular}{lcccc}
~~Channel   & Index & $\chi_J^{\sigma_i}$;~$\chi_j^c$ & $M$ & Mixed~~ \\
        &   &$[i; ~j]$ &  \\[2ex]
$(\phi \phi)^1 (2040)$     & 1  & [1;~1]   & $2022$ &  \\[2ex]
$(\phi \phi)^8$    & 2  & [1;~2]   & $2211$ &  \\[2ex]
$(ss)^*(\bar{s}\bar{s})^*$  & 3  & [1;~3]   & $2204$ & \\[2ex]
$K_1$   & 4  & [1;~5]   & $2143$ & \\
             & 5  & [1;~6]   & $2120$ & $2111$ \\[2ex]
$K_2$  & 6  & [1;~7]   & $2120$ & \\
             & 7  & [1;~8]   & $2143$ & $2111$ \\[2ex]
$K_3$  & 8  & [1;~10]   & $2203$ & \\[2ex]
$K_4$  & 9  & [1;~12]   & $2203$ & \\[2ex]
\multicolumn{4}{c}{Complete coupled-channels:} & $2012$ \\
\multicolumn{4}{c}{} & $E_B=-10$
\end{tabular}
\end{ruledtabular}
\end{table}

\begin{figure}[!t]
\includegraphics[clip, trim={3.0cm 1.7cm 3.0cm 0.8cm}, width=0.45\textwidth]{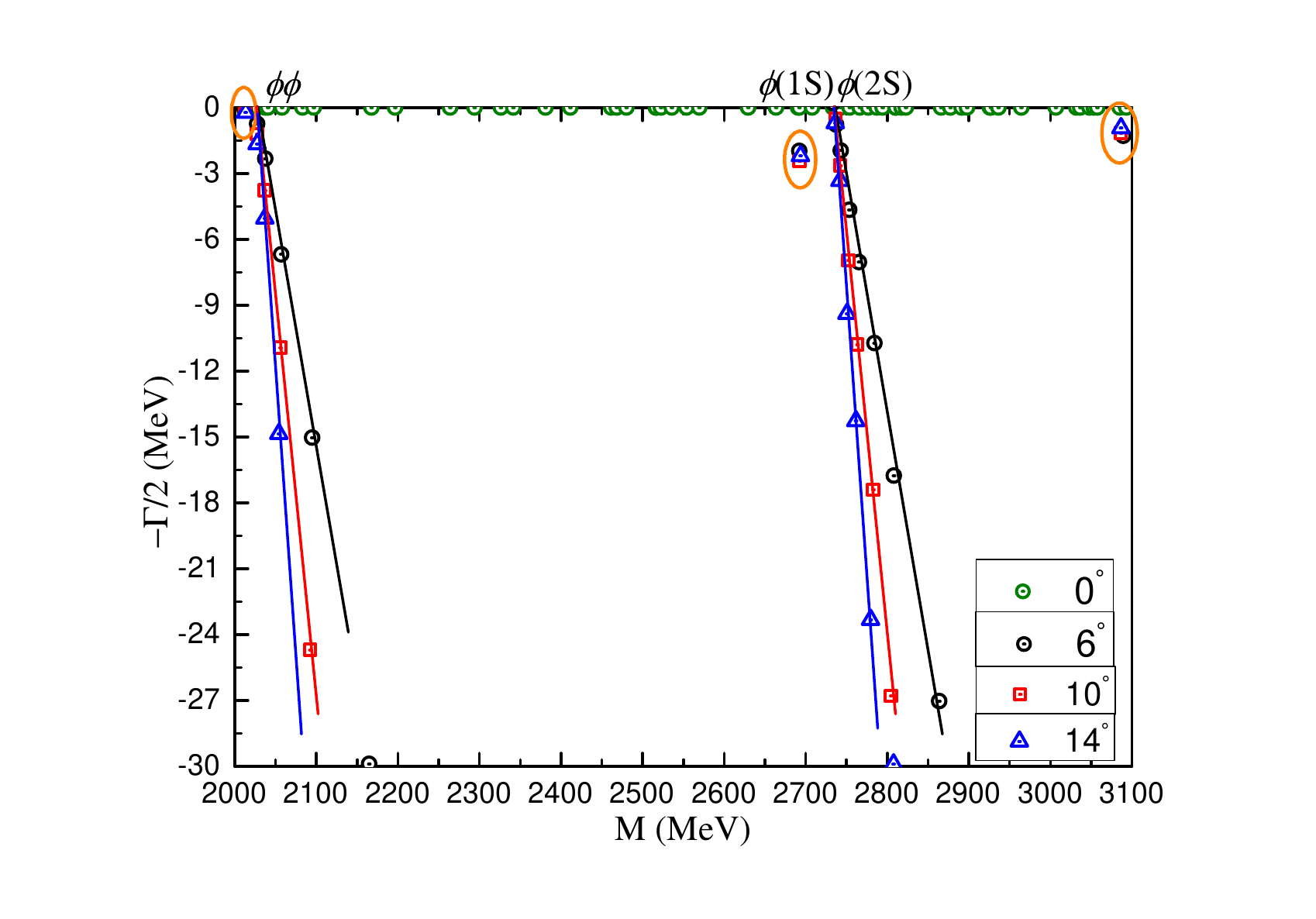}
\caption{\label{PP3} The fully coupled-channels calculation of $ss\bar{s}\bar{s}$ tetraquark system with $J^P=2^+$ quantum numbers.}
\end{figure}

\begin{table}[!t]
\caption{\label{GresultR3} Compositeness of exotic states obtained in a complete coupled-channel calculation in the $2^+$ state of $ss\bar{s}\bar{s}$ tetraquark. Results are similarly organized as those in Table~\ref{GresultR1}.}
\begin{ruledtabular}
\begin{tabular}{rccc}
Exotic state       & \multicolumn{3}{c}{Structure} \\[2ex]
$2012-i0$   & \multicolumn{3}{c}{$\mu=0$, $E_B=-10$} \\
  & \multicolumn{3}{c}{$r_{s \bar{s}}:1.15$;\,\,\,\,\,$r_{\bar{s}\bar{s}}:1.24$;\,\,\,\,\,$r_{ss}:1.24$} \\
$Set$ I: & \multicolumn{3}{c}{$S$: 6.1\%;\, $H$: 0\%;\, $Di$: 44.4\%;\, $K$: 49.5\%}\\
$Set$ II: & \multicolumn{3}{c}{$S$: 11.4\%;\, $H$: 0\%;\, $Di$: 11.7\%;\, $K$: 76.9\%}\\[2ex]
$2693-i4.9$   & \multicolumn{3}{c}{$\mu=0$} \\
  & \multicolumn{3}{c}{$r_{s \bar{s}}:1.23$;\,\,\,\,\,$r_{\bar{s}\bar{s}}:1.06$;\,\,\,\,\,$r_{ss}:1.06$} \\
$Set$ I: & \multicolumn{3}{c}{$S$: 0.8\%;\, $H$: 0\%;\, $Di$: 47.3\%;\, $K$: 51.9\%}\\
$Set$ II: & \multicolumn{3}{c}{$S$: 1.1\%;\, $H$: 0\%;\, $Di$: 36.9\%;\, $K$: 62\%}\\[2ex]
$3087-i2.3$   & \multicolumn{3}{c}{$\mu=0$} \\
  & \multicolumn{3}{c}{$r_{s \bar{s}}:1.66$;\,\,\,\,\,$r_{\bar{s}\bar{s}}:1.41$;\,\,\,\,\,$r_{ss}:1.41$} \\
$Set$ I: & \multicolumn{3}{c}{$S$: 0.1\%;\, $H$: 0\%;\, $Di$: 11.9\%;\, $K$: 88\%}\\
$Set$ II: & \multicolumn{3}{c}{$S$: 0.1\%;\, $H$: 0\%;\, $Di$: 23.9\%;\, $K$: 76\%}\\
\end{tabular}
\end{ruledtabular}
\end{table}

{\bf The $\bm{J^P=2^+}$ sector:}
The highest-spin configuration of the $ss\bar{s}\bar{s}$ system corresponds to the $J^P=2^+$ sector. As summarized in Table~\ref{GresultCC3}, a total of nine channels and their couplings are included in our analysis. No bound state is found in any single-channel calculation. The lowest-lying configuration corresponds to the $\phi\phi$ scattering state at its theoretical threshold of $2.02$ GeV, while the remaining eight channels populate the energy interval $2.12-2.21$ GeV. When all channels are fully coupled, however, a shallow bound state emerges with a theoretical mass of $2012$ MeV and a binding energy of $E_B=-10$ MeV. Applying the same mass-shift prescription used in the $J^P=0^+$ case, the modified mass of this bound state is estimated to be $2030$ MeV.

The results of the complete coupled-channel calculation using the Complex Scaling Method are shown in Fig.~\ref{PP3}. The scattering continua associated with the $\phi\phi$ and $\phi(1S)\phi(2S)$ channels are clearly identified in the mass range $2.0-3.1$ GeV. In addition, three isolated poles that remain stable under variations of the rotation angle $\theta$ are observed. Their complex energies are $M-i\Gamma = 2012 - i\,0$ MeV, $2693 - i\,4.9$ MeV, and $3087 - i\,2.3$ MeV, respectively. The first pole corresponds to the bound state already identified in the real-axis analysis, while the latter two represent resonance states. Notably, the resonance at $2.69$ GeV is close to the $X(2714)$ structure proposed in Ref.~\cite{Ma:2024vsi}.

The internal properties of these three states are summarized in Table~\ref{GresultR3}. Several common features emerge. First, all three states have vanishing magnetic moments. Second, they exhibit relatively loose spatial structures, with root-mean-square radii of approximately $1.2$ fm for the bound state and the lower resonance, increasing to about $1.4$ fm for the higher resonance at $3.09$ GeV. Third, a strong coupling between diquark-antidiquark and $K$-type configurations is observed in all three cases, indicating a pronounced compact multiquark character. These exotic states, characterized by nontrivial color structures, are therefore expected to be most readily identified through analyses of the $\phi\phi$ decay channel.


\begin{table}[!t]
\caption{\label{GresultCC4} Lowest-lying $ssss\bar{s}$ pentaquark states with $J^P=\frac{1}{2}^-$ calculated within the real range formulation of the chiral quark model. Results are similarly organized as those in Table~\ref{GresultCC1} (unit: MeV).}
\begin{ruledtabular}
\begin{tabular}{lcccc}
~~Channel   & Index & $\chi_J^{\sigma_i}$;~$\chi_j^c$ & $M$ & Mixed~~ \\
        &   &$[i; ~j]$ &  \\[2ex]
$(\Omega \phi)^1 (2692)$  & 1  & [1;~13]  & $2645$ & \\[2ex]
$(\Omega \phi)^8$          & 2  & [5;~14]  & $2783$ & \\
                                         & 3  & [4;~15]   & $2772$ &   \\
                                         & 4  & [3;~14]   & $2867$ &   \\
                                         & 5  & [2;~15]   & $2866$ &   \\
                                         & 6  & [1;~15]   & $3397$ & $2711$  \\[2ex]
$(ss)(ss)^*\bar{s}$          & 7  & [7;~16]  & $2844$ & \\
$(ss)^*(ss)^*\bar{s}$      & 8  & [9;~17]   & $2914$ &  \\
$(ss)^*(ss)^*\bar{s}$      & 9  & [8;~17]   & $2887$ & $2710$ \\[2ex]
\multicolumn{4}{c}{Complete coupled-channels:} & $2623$ \\
\multicolumn{4}{c}{} & $E_B=-22$
\end{tabular}
\end{ruledtabular}
\end{table}

\begin{table}[!t]
\caption{\label{GresultR4} Compositeness of exotic states obtained in a complete coupled-channel calculation in the $\frac{1}{2}^-$ state of $ssss\bar{s}$ pentaquark. Results are similarly organized as those in Table~\ref{GresultR1}.}
\begin{ruledtabular}
\begin{tabular}{rcc}
Exotic state       & \multicolumn{2}{c}{Structure} \\[2ex]
$2623-i0$   & \multicolumn{2}{c}{$\mu=-0.939$, $E_B=-22$} \\
  & \multicolumn{2}{c}{$r_{s \bar{s}}:1.24$;\,\,\,\,\,$r_{ss}:1.10$} \\
$Set$ I: & \multicolumn{2}{c}{$S$: 88.4\%;\, $H$: 10.7\%;\, $Di$: 0.9\%}\\
$Set$ II: & \multicolumn{2}{c}{$S$: 87.5\%;\, $H$: 11.4\%;\, $Di$: 1.1\%}\\ [2ex]
$2788-i0.2$   & \multicolumn{2}{c}{$\mu=-0.939$} \\
  & \multicolumn{2}{c}{$r_{s \bar{s}}:0.72$;\,\,\,\,\,$r_{ss}:0.68$} \\
$Set$ I: & \multicolumn{2}{c}{$S$: 0\%;\, $H$: 0\%;\, $Di$: 100\%}\\
$Set$ II: & \multicolumn{2}{c}{$S$: 0\%;\, $H$: 0\%;\, $Di$: 100\%}\\ [2ex]
$3214-i0.4$   & \multicolumn{2}{c}{$\mu=-0.936$} \\
  & \multicolumn{2}{c}{$r_{s \bar{s}}:1.31$;\,\,\,\,\,$r_{ss}:1.31$} \\
$Set$ I: & \multicolumn{2}{c}{$S$: 44.8\%;\, $H$: 41.4\%;\, $Di$: 13.8\%}\\
$Set$ II: & \multicolumn{2}{c}{$S$: 60.9\%;\, $H$: 23.7\%;\, $Di$: 15.4\%}\\
\end{tabular}
\end{ruledtabular}
\end{table}


\subsection{Fully strange pentaquarks}

{\bf The $\bm{J^P=\frac{1}{2}^-}$ sector:}
The results for the $ssss\bar s$ pentaquark in this quantum channel are summarized in Table~\ref{GresultCC4}. In the baryon-meson sector, the color-singlet $\Omega\phi$ configuration is unbound, with its lowest mass coinciding with the theoretical threshold at $2645$ MeV. In contrast, five hidden-color baryon-meson channels are obtained predominantly in the energy range $2.77-2.87$ GeV, with an additional channel appearing at a higher mass of $3.39$ GeV. Three diquark-diquark-antiquark configurations are also included, yielding masses clustered around $2.85$ GeV. When channel-coupling effects are taken into account separately within the exotic color structures, the corresponding lowest masses are reduced to approximately $2.71$ GeV. More significantly, a bound state emerges in the fully coupled-channel calculation. Its theoretical mass is $2623$ MeV, corresponding to a binding energy of $E_B=-22$ MeV. After shifting the mass relative to the experimental $\Omega\phi$ threshold, the modified mass of this bound state is $2670$ MeV.

The complete coupled-channel analysis using CSM, with the rotation angle $\theta$ varied from $0^{\circ}$ to $14^{\circ}$, is presented in Fig.~\ref{PP4}. Three scattering continua are clearly identified, corresponding to the $\Omega(1S)\phi(1S)$ ground state and the first radial excitations 
$\Omega(2S)\phi(1S)$ and $\Omega(1S)\phi(2S)$. In addition, three poles that remain stable against variations of $\theta$ are observed in the complex-energy plane. The lowest pole corresponds to the bound state at $2.62$ GeV, while the other two correspond to resonance states with complex energies 
$M-i\Gamma = 2788 - i\,0.2$ MeV and $3214-i\,0.4$ MeV, respectively.

The structural properties of the bound and resonant states are listed in Table~\ref{GresultR4}. All three states exhibit similar magnetic moments of approximately $-0.94\mu_N$. The bound state and the higher-mass resonance display relatively loose spatial structures, with root-mean-square radii of about $1.2$ fm. In contrast, the lower resonance at $2.79$ GeV is considerably more compact, with a characteristic size of roughly $0.7$ fm. This distinction is also reflected in their dominant configuration components. The higher resonance exhibits strong mixing among color-singlet, hidden-color, and diquark-diquark-antiquark channels, while the bound state contains sizable contributions from both color-singlet and color-octet baryon-meson configurations. By comparison, the lower resonance is found to be almost entirely dominated by a diquark-diquark-antiquark configuration.

Overall, the $J^P=\frac{1}{2}^-$ $ssss\bar s$ pentaquark sector exhibits a rich internal color structure, and the predicted bound and resonant states provide clear targets for future high-energy experimental searches.


\begin{table}[!t]
\caption{\label{GresultCC5} Lowest-lying $ssss\bar{s}$ pentaquark states with $J^P=\frac{3}{2}^-$ calculated within the real range formulation of the chiral quark model. Results are similarly organized as those in Table~\ref{GresultCC1} (unit: MeV).}
\begin{ruledtabular}
\begin{tabular}{lcccc}
~~Channel   & Index & $\chi_J^{\sigma_i}$;~$\chi_j^c$ & $M$ & Mixed~~ \\
        &   &$[i; ~j]$ &  \\[2ex]
$(\Omega \eta')^1 (2630)$   & 1  & [2;~13]  & $2462$ & \\
$(\Omega \phi)^1 (2692)$  & 2  & [1;~13]   & $2645$ & $2462$  \\[2ex]
$(\Omega \eta')^8$       & 3  & [4;~14]   & $2651$ &  \\
$(\Omega \eta')^8$       & 4  & [3;~15]  & $2641$ & \\
$(\Omega \phi)^8$       & 5  & [2;~15]  & $2900$ & \\
$(\Omega \phi)^8$   & 6  & [1;~15]   & $2704$ & $2633$ \\[2ex]
$(ss)(ss)^*\bar{s}$    & 7  & [5;~16]  & $2743$ & \\
$(ss)^*(ss)^*\bar{s}$    & 8  & [6;~17]   & $2790$ &  \\
$(ss)^*(ss)^*\bar{s}$    & 9  & [7;~17]   & $2745$ & $2734$  \\[2ex]  
\multicolumn{4}{c}{Complete coupled-channels:} & $2462$
\end{tabular}
\end{ruledtabular}
\end{table}

\begin{figure}[!t]
\includegraphics[clip, trim={3.0cm 1.7cm 3.0cm 0.8cm}, width=0.45\textwidth]{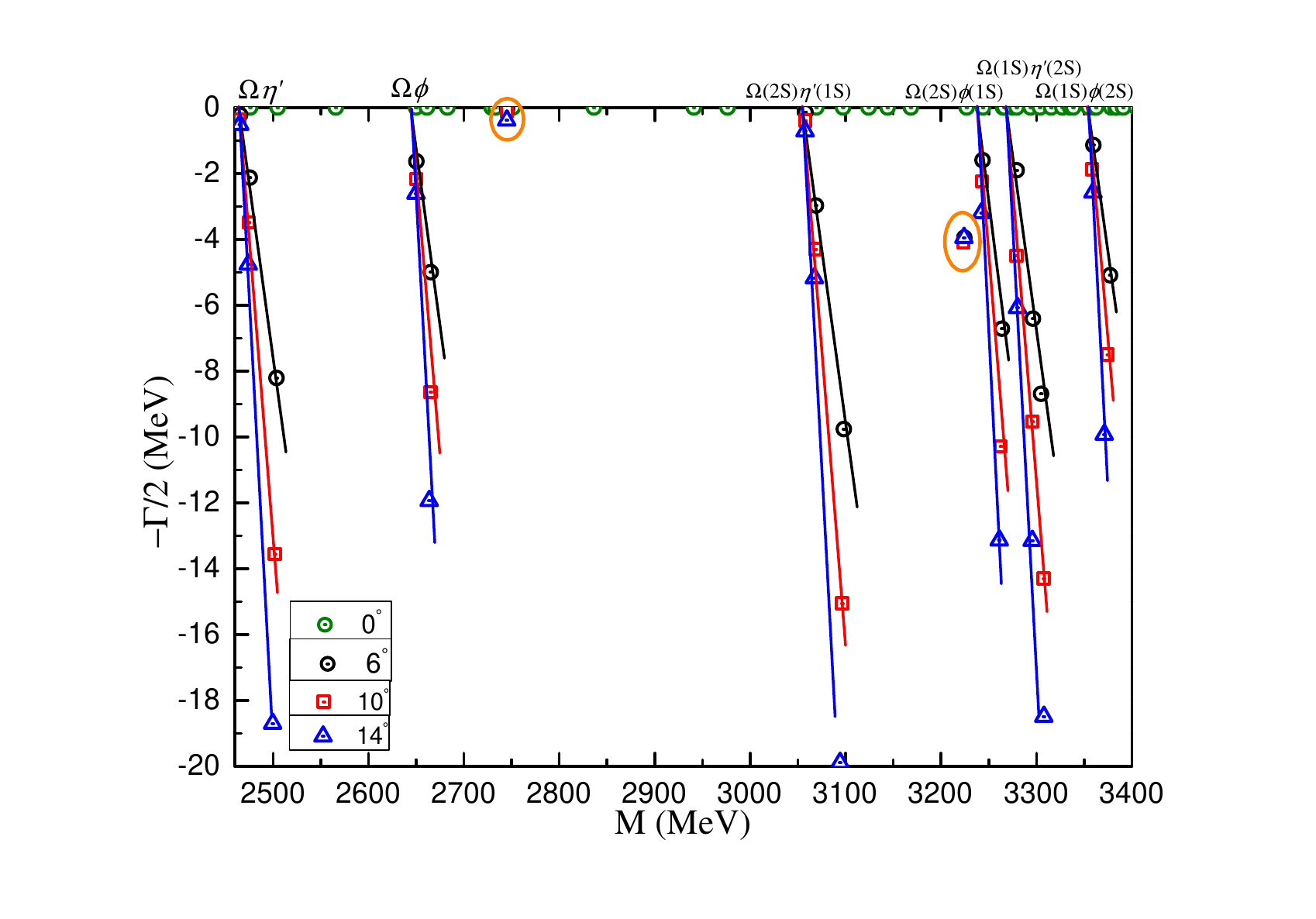}
\caption{\label{PP5} The fully coupled-channels calculation of $ssss\bar{s}$ pentaquark system with $J^P=\frac{3}{2}^-$ quantum numbers.}
\end{figure}

\begin{table}[!t]
\caption{\label{GresultR5} Compositeness of exotic resonances obtained in a complete coupled-channel calculation in the $\frac{3}{2}^-$ state of $ssss\bar{s}$ pentaquark. Results are similarly organized as those in Table~\ref{GresultR1}.}
\begin{ruledtabular}
\begin{tabular}{rcc}
Resonance       & \multicolumn{2}{c}{Structure} \\[2ex]
$2745-i0.2$   & \multicolumn{2}{c}{$\mu=-2.369$} \\
  & \multicolumn{2}{c}{$r_{s \bar{s}}:0.72$;\,\,\,\,\,$r_{ss}:0.67$} \\
$Set$ I: & \multicolumn{2}{c}{$S$: 0\%;\, $H$: 0\%;\, $Di$: 100\%}\\
$Set$ II: & \multicolumn{2}{c}{$S$: 0\%;\, $H$: 0\%;\, $Di$: 100\%}\\ [2ex]
$3223-i8.2$   & \multicolumn{2}{c}{$\mu=-1.483$} \\
  & \multicolumn{2}{c}{$r_{s \bar{s}}:1.33$;\,\,\,\,\,$r_{ss}:1.30$} \\
$Set$ I: & \multicolumn{2}{c}{$S$: 50.5\%;\, $H$: 48.2\%;\, $Di$: 1.3\%}\\
$Set$ II: & \multicolumn{2}{c}{$S$: 49.6\%;\, $H$: 46.2\%;\, $Di$: 4.2\%}
\end{tabular}
\end{ruledtabular}
\end{table}

{\bf The $\bm{J^P=\frac{3}{2}^-}$ sector:}
As summarized in Table~\ref{GresultCC5}, nine channels are investigated in this sector. Two baryon-meson configurations, $\Omega\eta'$ and $\Omega\phi$, are included in both color-singlet and hidden-color channels. All of these configurations are found to be unbound in single-channel calculations. The lowest masses of the color-singlet $\Omega\eta'$ and $\Omega\phi$ channels coincide with their respective theoretical thresholds at $2462$ MeV and $2645$ MeV. The hidden-color baryon-meson channels are generally located around $2.65$ GeV, with the exception of one $(\Omega\phi)^8$ configuration appearing at a higher mass of approximately $2.9$ GeV. In addition, three diquark-diquark-antiquark channels are considered, with masses clustered near $2.75$ GeV. Both partially and fully coupled-channel calculations performed in the real-energy framework indicate weak coupling effects, and no bound state is obtained in this sector.

A complete coupled-channel analysis using CSM reveals the presence of two resonance states. The distribution of complex eigenenergies is shown in Fig.~\ref{PP5}. In addition to six scattering continua corresponding to the $\Omega\eta'$, $\Omega\phi$, $\Omega(2S)\eta'(1S)$, $\Omega(2S)\phi(1S)$, $\Omega(1S)\eta'(2S)$, and $\Omega(1S)\phi(2S)$ channels, two poles that remain stable against variations of the rotation angle $\theta$ are identified. These resonances are located at complex energies $M-i\Gamma = 2745-i\,0.2$ MeV and $3223-i\,8.2$ MeV, respectively.

The properties of these resonant states are listed in Table~\ref{GresultR5}. The lower resonance at $2.74$ GeV has a magnetic moment of $-2.37,\mu_N$ and exhibits a compact $ssss\bar{s}$ pentaquark structure, with a characteristic size of approximately $0.7$ fm. Similar to the $2.79$ GeV resonance in the $J^P=\tfrac{1}{2}^-$ sector, this state is almost entirely dominated by a diquark-diquark-antiquark configuration. In contrast, the higher resonance at $3.22$ GeV has a magnetic moment of $-1.48,\mu_N$ and displays a relatively loose spatial structure, with a root-mean-square radius of about $1.3$ fm. In this case, strong coupling between color-singlet and hidden-color baryon-meson channels is observed. For both resonant states, the $\Omega\phi$ channel is expected to be the dominant strong-decay mode.


\begin{table}[!t]
\caption{\label{GresultCC6} Lowest-lying $ssss\bar{s}$ pentaquark states with $J^P=\frac{5}{2}^-$ calculated within the real range formulation of the chiral quark model. Results are similarly organized as those in Table~\ref{GresultCC1} (unit: MeV).}
\begin{ruledtabular}
\begin{tabular}{lcccc}
~~Channel   & Index & $\chi_J^{\sigma_i}$;~$\chi_j^c$ & $M$ & Mixed~~ \\
        &   &$[i; ~j]$ &  \\[2ex]
$(\Omega \phi)^1 (2692)$  & 1  & [1;~13]   & $2645$ & \\[2ex]
$(\Omega \phi)^8$             & 2  & [1;~15]   & $3193$ & \\[2ex]
$(ss)^*(ss)^*\bar{s}$         & 3  & [1;~17]   & $2726$ & \\[2ex]
\multicolumn{4}{c}{Complete coupled-channels:} & $2645$
\end{tabular}
\end{ruledtabular}
\end{table}

\begin{figure}[!t]
\includegraphics[clip, trim={3.0cm 1.7cm 3.0cm 0.8cm}, width=0.45\textwidth]{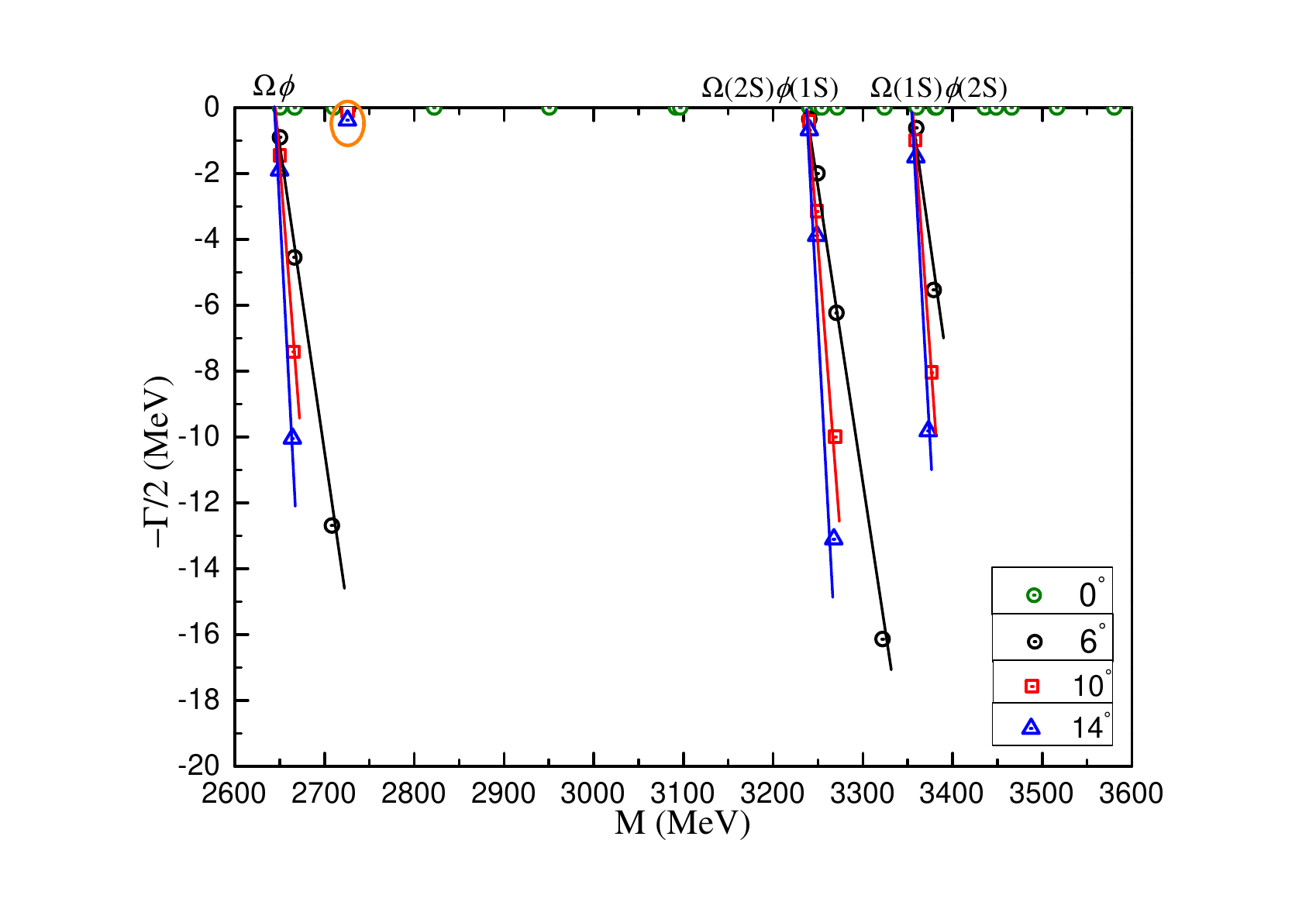}
\caption{\label{PP6} The fully coupled-channels calculation of $ssss\bar{s}$ pentaquark system with $J^P=\frac{5}{2}^-$ quantum numbers.}
\end{figure}

\begin{table}[!t]
\caption{\label{GresultR6} Compositeness of exotic resonance obtained in a complete coupled-channel calculation in the $\frac{5}{2}^-$ state of $ssss\bar{s}$ pentaquark. Results are similarly organized as those in Table~\ref{GresultR1}.}
\begin{ruledtabular}
\begin{tabular}{rcc}
Resonance       & \multicolumn{2}{c}{Structure} \\[2ex]
$2726-i0.8$   & \multicolumn{2}{c}{$\mu=-1.692$} \\
  & \multicolumn{2}{c}{$r_{s \bar{s}}:0.70$;\,\,\,\,\,$r_{ss}:0.67$} \\
$Set$ I: & \multicolumn{2}{c}{$S$: 0\%;\, $H$: 0\%;\, $Di$: 100\%}\\
$Set$ II: & \multicolumn{2}{c}{$S$: 0\%;\, $H$: 0\%;\, $Di$: 100\%}
\end{tabular}
\end{ruledtabular}
\end{table}

{\bf The $\bm{J^P=\frac{5}{2}^-}$ sector:}
For the highest-spin configuration of the fully strange pentaquark, we consider three channels, as listed in Table~\ref{GresultCC6}. Specifically, the lowest masses of the $\Omega \phi$ system in the singlet- and octet-color configurations are $2645$ MeV and $3193$ MeV, respectively. The $(ss)^*(ss)^*\bar{s}$ channel appears at $2726$ MeV. As a result, no bound state is found in this sector, and this conclusion remains valid in a complete coupled-channel analysis.

Figure~\ref{PP6} shows that, within the energy range $2.6-3.6$ GeV, the ground state of $\Omega \phi$ and the first radial excitations $\Omega(2S)\phi(1S)$ and $\Omega(1S)\phi(2S)$ are clearly identified. In addition, one stable resonance pole is observed, with a complex energy of $M-i\Gamma=2726 - i\,0.8$ MeV.

A summary of the resonant state is provided in Table~\ref{GresultR6}. Notably, the magnetic moment of this exotic state is $-1.69,\mu_N$. Its spatial structure and internal composition are similar to the lower resonances found in the $J^P = 1/2^-$ and $3/2^-$ sectors. The state exhibits a compact pentaquark configuration with an approximate size of $0.7$ fm and represents an exotic color resonance, consisting purely of the $(ss)^*(ss)^*\bar{s}$ component. These predictions could be tested in future high-energy experiments.


\begin{table}[!t]
\caption{\label{GresultCCT} Summary of exotic states structures found in the fully strange tetra- and penta-quark systems. The first column shows the total spin and parity of each singularity. The second column refers to the theoretical pole with notation: $E=M-i\Gamma$, if possible, the binding energy ($E_B$) of bound state is labeled in parentheses (unit: MeV). Size ($r$, unit: fm) and magnetic moment ($\mu$, unit: $\mu_N$) of state are presented in the last column.}
\begin{ruledtabular}
\begin{tabular}{lcc}
~ $J^P$ & Theoretical pole   & Structure~~ \\
                   & $E=M-i\Gamma$ $(E_B)$   & $r,\,\mu$ \\
\hline
\multicolumn{3}{c}{$ss\bar{s}\bar{s}$ tetraquarks}\\
~~$0^+$  & $1645-i0$ $(-11)$      & $1.13\sim1.21$, $0$~~  \\
                & $2067-i0.1$   & $0.73\sim1.21$, $0$~~  \\
                & $2506-i0.4$   & $1.19\sim1.43$, $0$~~  \\[2ex]
~~$1^+$  & $2262-i0.6$   & $0.73\sim0.79$, $0$~~ \\
                & $2638-i11.3$   & $1.31\sim1.50$, $0$~~  \\[2ex]
~~$2^+$  & $2012-i0$ $(-10)$  & $1.15\sim1.24$, $0$~~ \\
                & $2693-i4.9$   & $1.06\sim1.23$, $0$~~ \\
                & $3087-i2.3$   & $1.41\sim1.66$, $0$~~ \\
\hline
\multicolumn{3}{c}{$ssss\bar{s}$ pentaquarks}\\
~~$\frac{1}{2}^-$    & $2623-i0$ $(-22)$  & $1.10\sim1.24$, $-0.939$~~  \\
                                 & $2788-i0.2$   & $0.68\sim0.72$, $-0.939$~~  \\
                                 & $3214-i0.4$   & $1.31$, $-0.936$~~  \\[2ex]
~~$\frac{3}{2}^-$   & $2745-i0.2$  & $0.67\sim0.72$, $-2.369$~~  \\
                                 & $3223-i8.2$   & $1.30\sim1.33$, $-1.483$~~  \\[2ex]
~~$\frac{5}{2}^-$   & $2726-i0.8$   & $0.67\sim0.70$, $-1.692$~~
\end{tabular}
\end{ruledtabular}
\end{table}

\section{Summary}
\label{sec:summary}

In this work, we have carried out a systematic investigation of fully strange tetra- and penta-quark systems within a chiral constituent quark model, motivated in particular by the recent observation of the $X(2300)$ resonance. Fully strange multiquark states provide a unique environment to explore nonperturbative QCD dynamics, as they are in the intersection between the light and heavy quark sectors. The main goal of this study has been to clarify the spectrum, internal structure, and nature of low-lying $S$-wave $ss\bar s\bar s$ and $ssss\bar s$ systems.

Our analysis employs the Gaussian Expansion Method combined with the Complex Scaling Method, allowing for a unified and consistent treatment of bound states, resonances, and scattering states. A key feature of the present work is the complete inclusion of all relevant $S$-wave configurations and color structures. For tetraquarks, meson-meson, diquark-antidiquark, and K-type configurations, including both color-singlet and hidden-color channels, are considered on equal footing. For pentaquarks, both baryon-meson and diquark-diquark-antiquark arrangements are simultaneously incorporated. This comprehensive framework enables us to reliably identify genuine multiquark poles and disentangle them from continuum effects.

In the fully strange tetraquark sector, several weakly bound states and narrow resonances are obtained in the $J^P=0^+$, $1^+$, and $2^+$ channels. In particular, a compact tetraquark state with $J^P=1^+$ is found near $2.3$ GeV, whose mass, width, and internal structure provide a natural interpretation of the experimentally observed $X(2300)$. Our results support an $S$-wave fully strange tetraquark assignment for this state, characterized by a dominant hidden-color component and a compact spatial size. Additional narrow resonances are predicted at higher energies, some of which are dominated by exotic K-type configurations that are rarely considered in phenomenological studies.

In the fully strange pentaquark sector, we predict several bound and resonant states with quantum numbers $J^P=1/2^-$, $3/2^-$, and $5/2^-$ in the mass region between approximately $2.6$ and $3.2$ GeV. These states exhibit a rich internal structure, with significant contributions from diquark-diquark-antiquark configurations, indicating that genuinely exotic color correlations play an essential role in stabilizing fully strange pentaquarks. The narrow widths of some of these states suggest that they may be accessible in future high-statistics experiments.

Beyond the mass spectrum, we have analyzed the internal properties of the predicted exotic states through their root-mean-square radii, magnetic moments, and wave-function compositions. These observables provide clear evidence that channel coupling effects and hidden-color configurations are crucial for the formation of bound and resonant multiquark states. In particular, states dominated by K-type tetraquark structures or diquark-diquark-antiquark pentaquark configurations tend to be more compact and less molecular in nature.

Table~\ref{GresultCCT} summarizes our results on the mentioned exotic bound states and resonances, which includes the complex energy, $M-i\,\Gamma$, the binding energy for bound states, $E_B$, as well as their root-mean-square radii, $r$, and magnetic moments, $\mu$.

Finally, we have identified promising two-body strong decay channels for several tetra- and penta-quark states, which can serve as {\it golden modes} for future experimental searches. The present results therefore provide not only a coherent theoretical interpretation of existing experimental observations, such as the $X(2300)$, but also concrete predictions that can be tested in forthcoming experiments at BESIII and other facilities.

In summary, this study highlights the importance of exotic color structures and complete channel coupling in understanding fully strange multiquark systems. We expect that further experimental progress in the strange sector will play a decisive role in establishing the existence and nature of these exotic hadrons, thereby deepening our understanding of multiquark dynamics in QCD.


\begin{acknowledgments}
Work partially financed by National Natural Science Foundation of China under Grant No. 12305093; Ministerio Espa\~nol de Ciencia e Innovaci\'on under grant No. PID2022-140440NB-C22; Junta de Andaluc\'ia under contract No. FQM-370 as well as PCI+D+i under the title: ``Tecnolog\'\i as avanzadas para la exploraci\'on del universo y sus componentes" (Code AST22-0001).
\end{acknowledgments}


\bibliography{FSTP}

\end{document}